\documentclass[useAMS,usenatbib]{mnras}

\usepackage{aas_macros}
\usepackage{gensymb}
\usepackage{amsmath}
\usepackage{amssymb}
\usepackage{float}
\usepackage{graphicx}
\usepackage{lscape}
\usepackage{url}
\usepackage{natbib}
\usepackage{placeins}
\usepackage{color}
\usepackage{wasysym}
\usepackage{hyperref}
\usepackage{scrextend}
\usepackage{quotes}

\newcommand{\nhii}{\ensuremath{n_{{\mbox{\small{H}\sc{ii}}}}}}
\newcommand{\nhi}{\ensuremath{n_{\mbox{\small{H}\sc{i}}}}}
\newcommand{\nH}{\ensuremath{n_{{\mbox{\small{H}}}}}}

\hypersetup{draft}
\begin{document}

\title[Effects of Radiation Pressure on HD~209458b]{Effects of Radiation Pressure on the Evaporative Wind of HD~209458b}

\author[A. Debrecht et al.]{Alex Debrecht$^{1}$\thanks{adebrech@ur.rochester.edu}, Jonathan Carroll-Nellenback$^{1}$, Adam Frank$^{1}$,
\newauthor
Eric G.~Blackman$^{1}$, Luca Fossati$^{2}$, John McCann$^{3}$, Ruth Murray-Clay$^{4}$ \\
$^{1}$Department of Physics and Astronomy, University of Rochester, Rochester NY 14627\\
$^{2}$Space Research Institute, Austrian Academy of Sciences, Schmiedlstrasse 6, A-8042 Graz, Austria\\
$^{3}$Department of Physics, University of California, Santa Barbara, Santa Barbara, CA, 93106\\
$^{4}$Physics and Astronomy Department, University of California, Santa Cruz, Santa Cruz, CA 95064\\
}

\date{}

\pagerange{\pageref{firstpage}--\pageref{lastpage}}
\maketitle
\label{firstpage}

\begin{abstract}
The role of radiation pressure in shaping exoplanet photoevaporation remains a topic of contention. Radiation pressure from the exoplanet's host star has been proposed as a mechanism to drive the escaping atmosphere into a "cometary" tail and explain the high velocities observed in systems where mass loss is occurring. In this paper we present results from high-resolution 3-D hydrodynamic simulations of a planet similar to HD~209458b. We self-consistently launch a wind flowing outward from the planet by calculating the ionization and heating resulting from incident high-energy radiation, and account for radiation pressure. We first present a simplified calculation, setting a limit on the Lyman-$\alpha$ flux required to drive the photo-evaporated planetary material to larger radii and line-of-sight velocities. We then present the results of our simulations, which confirm the limits determined by our analytic calculation. We thus demonstrate that, within the limits of our hydrodynamic simulation and for the Lyman-$\alpha$ fluxes expected for HD~209458, radiation pressure is unlikely to significantly affect photoevaporative winds or to explain the high velocities at which wind material is observed, though further possibilities remain to be investigated.
\end{abstract}

\begin{keywords}
hydrodynamics -- planet-star interactions -- planets and satellites: atmospheres -- planets and satellites: individual: HD 209458b
\end{keywords}

\section{Introduction}

Characterization of exoplanetary atmospheres is now a dynamic field of research. Much of the literature has focused on atmospheric structure, composition, and the detection of biosignatures, but the interaction of a planetary atmosphere with the circumplanetary environment remains an important area of inquiry, as it plays a key role in determining the long-term evolution of planetary atmospheres. In particular, atmospheric escape, caused by the absorption of stellar X-ray and extreme ultraviolet radiation by the atmosphere of close-in planets, is of particular interest, both theoretically \citep[e.g.][]{lammer2003, yelle2004, lecavelier04, tian05, garciamunoz07, murrayclay09, koskinen2013a, khodachenko17} and observationally \citep[e.g.][]{vidalmadjar03, fossati10a, lecavelier12, ehrenreich15, bourrier18}.

From the theoretical side, many physical processes can affect the evolution of evaporative planetary winds, including stellar and planetary magnetic fields \citep{matsakos15, villarreal18, owen14, khodachenko15} and time-dependent phenomena such as flares \citep{bisikalo2018}, coronal mass ejections \citep{cherenkov2017}, stellar variability \citep{lecavelier12, bisikalo16}, and atmospheric circulation \citep{teyssandier15}. 

Observationally, photoevaporative outflows have been directly studied in some detail for a handful of planets. These include HD~209458b \citep{vidalmadjar03}, HD~189733b \citep{lecavelier10, lecavelier12, bourrier13b}, GJ~436b \citep{kulow14, ehrenreich15}, and GJ~3470b \citep{bourrier18}. For many of these planets, asymmetric H{\sc i} absorption of the stellar Lyman-$\alpha$ emission line by material of planetary origin has been observed, with enhanced absorption seen in the blue wing when compared to the red wing. The high velocities of the absorbing material of up to $\pm 150$ km/s observed in these systems challenge models of exoplanet evaporation, which predict lower planetary wind speeds of order the planetary escape speed ($\sim30$ km/s).

Several processes have been suggested as the cause of this asymmetric high-velocity absorption, including charge exchange between the stellar and planetary wind \citep{holmstrom2008, ekenback10, tremblin13, kislyakova2014, bourrier16, christie16}, confinement by the stellar wind \citep{Schneiter2007, Schneiter2017, mccann18}, and acceleration of neutrals by radiation pressure from stellar Lyman-$\alpha$ emission \citep{lecavelier08, Schneiter2017, bourrier13, khodachenko17, cherenkov18}. The differing conclusions resulting from these studies show that it is not yet clear which if any of these processes can fully account for the observations. Here we address this question by focusing on one process: radiation pressure from the host star.

A number of studies have supported the idea that radiation pressure can change the photoevaporative flows. \citet{Schneiter2017}, for example, performed detailed simulations including both the star and the planet, finding that the planetary wind was driven back into a cometary tail. However, in these simulations both planet and star were modeled as isotropic sources, with the planetary wind launched from outside the planet's atmosphere at 3 planetary radii ($R_p$). In addition, radiation pressure from the star was modeled as a reduction in the stellar gravity throughout the grid. \citet{bourrier13} also evolved the wind from between the Roche lobe and the planet's surface isotropically. However, while they also reduced the stellar gravity to model radiation pressure, they included self-shielding from absorption of the Lyman-$\alpha$ photons. In addition, \citet{bourrier13} removed hydrogen particles from the simulation once they were ionized. This study also found that radiation pressure produces a cometary tail-like structure in the planetary wind, with much of the gas escaping the system.

On the other hand, \citet{khodachenko17} have recently presented 2-dimensional hydrodynamic simulations of a four-fluid model of HD~209458b, using a simplified plateau-like Lyman-$\alpha$ profile and a velocity-dependent absorption of Lyman-$\alpha$ photons by neutral hydrogen. The planetary wind is launched by heating from EUV radiation and interacts with the stellar wind through charge exchange. They find that radiation pressure does not significantly affect the planetary wind. \citet{cherenkov18} calculated the ionization and radiation pressure self-consistently in a 3D simulation at moderate resolution, using a hydrogen envelope to simulate HD~209458b. They find that the planetary wind would be affected by Lyman-$\alpha$ radiation only if it were two orders of magnitude greater than that predicted for HD~209458b.

Thus the question of the role of radiation pressure on exoplanetary photoevaporative flows remains quite open. In this paper we present results from high-resolution 3-D hydrodynamic simulations of a planet similar to HD~209458b, launching the wind self-consistently by calculating the ionizing radiation and accounting for radiation pressure. In section \ref{sec:meth} we present the computational method and parameters used in our simulations. In section \ref{sec:result} we show the results of our simulations. In section \ref{sec:disc}, we compare our results to the analytic calculation and the results of previous studies and suggest possible reasons for our lack of a significant cometary tail and high-velocity absorption signature. We conclude in section \ref{sec:conc}.

\section{Methods and Model} \label{sec:meth}

Our simulations were conducted with AstroBEAR\footnote{https://astrobear.pas.rochester.edu/} \citep{cunningham09,carroll13}, a massively parallelized adaptive mesh refinement (AMR) code that includes a variety of multiphysics solvers, such as self-gravity, heat conduction, magnetic resistivity, radiative transport, and ionization dynamics. The equations solved for these simulations are those of fluid dynamics in a rotating reference frame, including gravitational effects of both planet and star:
\begin{equation}
 \frac{\partial \rho}{\partial t} + \boldsymbol{\nabla} \cdot \rho \boldsymbol{v} = 0,
 \label{eq:Eu1}
\end{equation}
\begin{equation}
 \frac{\partial \rho \boldsymbol{v}}{\partial t} + \boldsymbol{\nabla} \cdot \left ( \rho \boldsymbol{v} \otimes \boldsymbol{v} \right )= - \boldsymbol{\nabla} p - \rho \boldsymbol{\nabla} \phi + \boldsymbol{f_R} + \boldsymbol{\mathcal{K}},
 \label{eq:Eu2}
\end{equation}
\begin{equation}
 \frac{\partial E}{\partial t} + \boldsymbol{\nabla} \cdot ((E + p) \boldsymbol{v}) = -\rho \boldsymbol{v} \cdot \boldsymbol{\nabla} \phi + \boldsymbol{f_R} \cdot \boldsymbol{v} + \mathcal{G} - \mathcal{L} + \mathcal{J},
 \label{eq:Eu3}
\end{equation}
where $\rho$ is the mass density, $\boldsymbol{v}$ is the fluid velocity, $p$ is the thermal pressure, $\phi$ is the gravitational potential, $\boldsymbol{f_R}$ combines the Coriolis and centrifugal forces so that $\boldsymbol{f_R} = \rho \left ( - 2 \boldsymbol{\Omega} \times \boldsymbol{v} - \boldsymbol{\Omega} \times \left ( \boldsymbol{\Omega} \times \boldsymbol{r} \right ) \right )$ (where $\boldsymbol{\Omega}$ is the orbital velocity), $E = p/(\gamma - 1) + \rho v^2/2$ is the combined internal and kinetic energies, $\mathcal{G}$ and $\mathcal{L}$ are respectively the heating and cooling rates, $\boldsymbol{\mathcal{K}}$ is the force due to Lyman-$\alpha$ radiation, and $\mathcal{J}$ is the kinetic energy added by the radiation pressure from Lyman-$\alpha$ radiation.

The simulation also tracked the advection, photoionization, and recombination of neutral and ionized hydrogen. We use the photon-conserving update scheme from \citet{krumholz07} to solve the following equations:
\begin{equation}
 \frac{\partial \nhi}{\partial t} + \boldsymbol{\nabla} \cdot (\nhi \boldsymbol{v}) = \mathcal{R} - \mathcal{I},
 \label{eq:Eu4}
\end{equation} 
\begin{equation}
 \frac{\partial \nhii}{\partial t} + \boldsymbol{\nabla} \cdot (\nhii \boldsymbol{v}) = \mathcal{I} - \mathcal{R},
\end{equation}
where $\nhi$ is the number density of neutral hydrogen, $\nhii$ is the number density of ionized hydrogen, and $\mathcal{R}$ and $\mathcal{I}$ are the recombination and ionization rates.

\subsection{Radiation transfer} \label{sec:rad_trans}

We perform the radiation transfer as in \citet{debrecht19}. In addition to the ionizing flux, we apply a monochromatic Lyman-$\alpha$ flux $F_{0,\alpha}$. We calculate its radiative transfer using the same method used for the ionizing flux, but with an absorption cross-section $\sigma_{\alpha} = 5.9\times10^{-14} \left(10^4 \mbox{ K}/T\right)^{1/2}$. Each absorbed Lyman-$\alpha$ photon deposits its entire momentum into the gas, so the (volumetric) force due to radiation pressure is 
\begin{equation}
    \boldsymbol{\mathcal{K}} = \sigma_{\alpha} \nhi F_{\alpha}(x) p_{\gamma} \hat{x}
\end{equation}
and the rate of kinetic energy deposition is
\begin{equation}
    \mathcal{J} = \boldsymbol{\mathcal{K}} \cdot \boldsymbol{v}
\end{equation}
where $p_{\gamma} = 10.2 \mbox{ eV}/c$ is the momentum per Lyman-$\alpha$ photon, $p_{0,x}$ is the initial momentum of the cell in the $x$ direction, $dt$ is the time step of the radiation calculation, and $\rho$ is the density of the cell. Note that the gas density does not change during the radiation calculation, and that our radiation is represented by a single frequency bin, thereby ignoring the Doppler shift of the absorption line. Consequences of this assumption are discussed further in section \ref{sec:cptoan}.

\subsection{Planet atmosphere model} \label{sec:atmo}

In our simulations, we have modeled the planet as a sphere of hydrogen in hydrostatic equilibrium as in \citet{debrecht19}, with a density profile of
\begin{equation}
 \rho_{atm}(r) = \rho_p \left [\frac{R_0 R_p}{R_0 - R_p} \left (\frac1r -\frac1{R_0}\right)\right]^{\frac1{\gamma-1}}
\end{equation}
and pressure profile of 
\begin{equation}
 P_{atm}(r) = \rho_p^{1-\gamma} c_{s,p}^2 \rho^\gamma,
\end{equation}
where $R_p$ is the radius of the planet and $R_0$ is the radius where a physical atmosphere would end. For completeness, we now describe the internal planetary boundaries used in the simulation to model the planet. $R_{ob}$, three cells inside $R_0$, is the radius at which the planetary profile is initially cut off and the ambient conditions begin. This is done to reduce numerical noise at the interface between the edge of the planet and the ambient material. $R_{ob} = 1.04 R_p$. $R_{ib} = R_0/(1 + (R_0-R_p)/(R_p e^{3(\gamma - 1)}))$ is the radius at which the planetary profile is reset at every time step in order to replenish the supply of atmospheric hydrogen flowing into the outer layers of the planet from its core. $R_{ib} = 0.68 R_p$. $R_{mask}$, set five cells within $R_{ib}$, is the radius at which the planetary profile is cut off to prevent the singularity at $r = 0$. $R_{mask} = 0.62 R_p$. As long as the radius from which the wind is launched (the $\tau = 1$ radius) does not reach $R_{ib}$, the structure of the wind will not be significantly affected by any of these choices \citep[see Appendix A of][]{murrayclay09}. In addition, the mass lost from the planet during the simulation, approximately $10^{15}$ g at $5\times10^9$ g/s, is only about 1\% of the total initial mass, suggesting that the structure of the layers does not have significant bearing on the evolution of the wind. $\rho_p$ is the density and $c_{s,p}^2$ is the sound speed at the planetary surface $R_p$.

\begin{figure}
\centering
\includegraphics[width=\columnwidth]{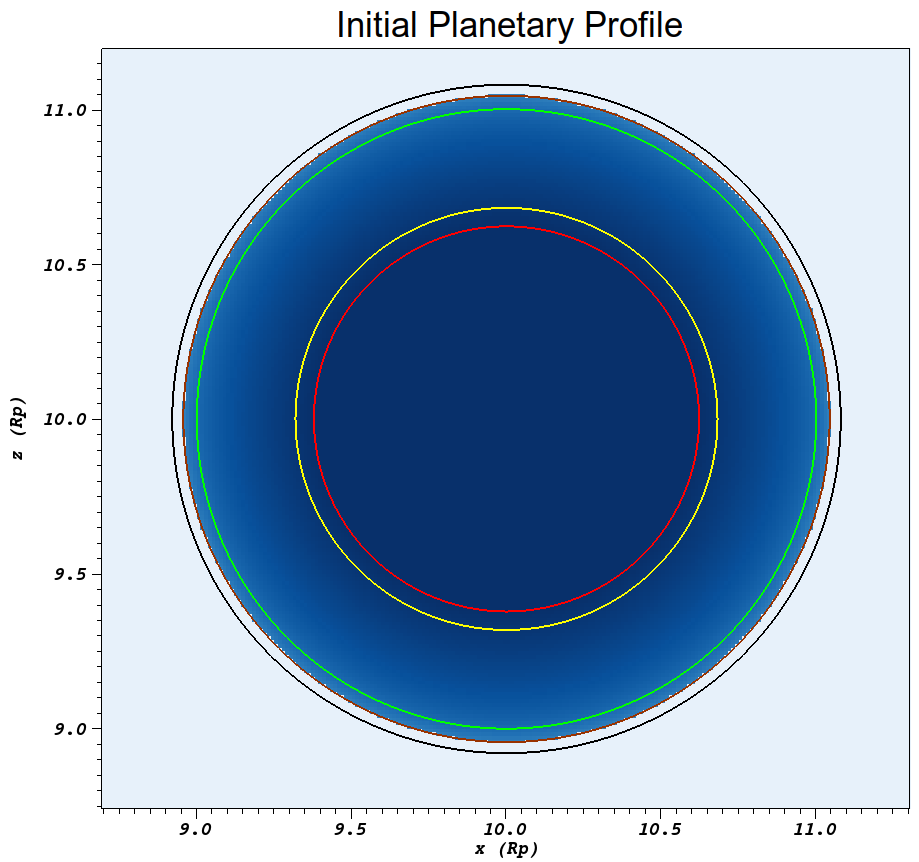}
\caption{The initial state of our planet, as described in section \ref{sec:atmo}. From inside to outside, the contours show: red: $R_{mask}$; yellow: $R_{ib}$; green: $R_p$; brown: $R_{ob}$; black: $R_0$.}
\label{fig:planet_profile}
\end{figure}

In addition, we now set up a static ambient pressure two orders of magnitude lower than the final value of the planet profile, rather than matching the pressure at the boundary. The complete specification of the initial conditions of the simulation is thus
\begin{equation}
 \rho(r) = 
 \begin{cases}
  \rho_{atm}(R_{mask}), & r < R_{mask} \\
  \rho_{atm}(r), & R_{mask} \leq r \leq R_{ob} \\
  \rho_{atm}(R_{ob}) \times 10^{-4}, & r > R_{ob}
 \end{cases},
\end{equation}
with corresponding pressures of
\begin{equation}
 P(r) = 
 \begin{cases}
  P_{atm}(R_{mask}), & r < R_{mask} \\
  P_{atm}(r), & R_{mask} \leq r \leq R_{ob} \\
  P_{atm}(R_{ob}) \times 10^{-2}, & r > R_{ob}
 \end{cases}.
\end{equation}
A sketch of the different layers is shown in figure \ref{fig:planet_profile}.

\subsection{Description of simulation}

The input parameters of the simulations were chosen to model a planet similar to HD~209458b. Our planet has a radius of $1.529 R_{J}$, chosen to be 1.1 times the measured broadband optical transit radius $R_{tr} = 1.39 R_{J}$ (based on the discussion in section 2 of \citet{murrayclay09} of the predicted wind launch radius), a mass of $0.73 M_{J}$, and a temperature of 3000 K. While this is higher than the predicted equilibrium temperature of 1500 K, it has been shown that the planet's surface temperature has a negligible effect on the wind structure for any temperature below the wind temperature at the launching surface \citep[][their Appendix A]{murrayclay09}, which is generally $\sim 10^4 \mbox{ K}$. The planet orbits the host star of mass $1.23 M_{\astrosun}$ and radius $1.19 R_{\astrosun}$ at a separation of $a = 0.047 \mbox{ AU}$. Planetary and stellar parameters were taken from \citet{stassun17}.

The Cartesian simulation domain ranges from $[-10,-10,-10] R_p$ to $[10,10,10] R_p$, with the planet centered at $[0,0,0]$. We apply outflow-only extrapolating boundary conditions at all boundaries, with the initial ambient conditions applied if the extrapolated conditions would result in inflow. The simulation has a base resolution of $105^3$ and 4 levels of additional refinement, giving an effective resolution of $1680^3$, which resolves the entire grid to $0.012 R_p$. The maximum resolution is forced in a sphere out to the termination radius $R_0$ of the hydrostatic atmosphere (defined further in Section \ref{sec:atmo}). The planetary radius is therefore resolved by 84 cells. We allow the mesh to evolve outside of the planet based on the density gradient. The stellar location is not included in these simulations; only its radiation and gravitational effects are considered. In order to isolate the effect of radiation pressure, we have not included a stellar wind in these simulations.

We have run three simulations, varying the Lyman-$\alpha$ flux by an order of magnitude between each. Our fiducial value of $4.1\times10^{14} \mbox{ phot cm}^{-2}\mbox{s}^{-1}$ was calculated from the value of the Lyman-$\alpha$ flux of HD~209458 measured by \citet{wood05}. The hydrogen density $\rho_p = 1.625\times10^{-15} \mbox{ g cm}^-3$ at the planetary radius $R_p$ was adjusted so that the surface of unit optical depth to ionizing radiation at 16 eV for the un-irradiated planet, along the substellar line, was near the planet radius $R_p$. 

For our base simulation, we evolved the planet wind for 4.58 days (1.3 planetary orbits) to a steady state without Lyman-$\alpha$ flux. After introducing the Lyman-$\alpha$ flux, we ran the simulations for 2.5 days (0.72 planetary orbits) for the low- and intermediate-flux runs and 3.4 days (0.97 planetary orbits) for the high-flux run, after which the low- and intermediate-flux simulations (runs 1 and 2 in table \ref{tab:runs}) had reached a steady state, by which we mean the flow had achieved a stable ionization front and wind morphology. The high-flux simulation (Run 3 in table \ref{tab:runs}), however, enters a cycle whereby the wind drives away from the planet, is disrupted by radiation pressure, and drives away from the planet again, repeating the cycle. This simulation was run for a time sufficient to see two repetitions of this cycle.

\setcounter{table}{0}

\begin{table*}
\centering

 \caption{Run parameters}
 \label{tab:runs}
 \begin{tabular}{l|c|c|c|c|c|}

  \hline

  & & No-Flux & Low-Flux & Int.-Flux & High-Flux \\

  \hline
  
  Planet Radius & $R_p\,(R_J)$ & \multicolumn{4}{c}{$1.529$} \\
  Planet Mass & $M_p\,(M_J)$ & \multicolumn{4}{c}{$0.73$} \\
  Planet Temperature & $T_p\,(\mbox{K})$ & \multicolumn{4}{c}{$3\times10^3$} \\
  Planet Surface Density & $\rho_p\,(\mbox{g cm}^{-3})$ & \multicolumn{4}{c}{$1.625\times10^{-15}$} \\
  Stellar Mass & $M_\star\,(M_{\astrosun})$ & \multicolumn{4}{c}{$1.23$} \\
  Stellar Radius & $R_\star\,(R_{\astrosun})$ & \multicolumn{4}{c}{$1.19$} \\
  Stellar Ionizing Flux & $F_{0,UV}\,(\mbox{phot cm}^{-2}\mbox{s}^{-1})$ & \multicolumn{4}{c}{$2\times10^{13}$} \\
  Stellar Lyman-$\alpha$ Flux & $F_{0,\alpha}\,(\mbox{phot cm}^{-2}\mbox{s}^{-1})$ & $0$ &  $4.1\times10^{13}$ & $4.1\times10^{14}$ & $4.1\times10^{15}$ \\
  Orbital Separation & $a$ (AU) & \multicolumn{4}{c}{$0.047$} \\
  Orbital Period & $P$ (days) & \multicolumn{4}{c}{$3.525$} \\
  Orbital Velocity & $\Omega$ (rad/day) & \multicolumn{4}{c}{$1.78$} \\
  Polytropic Index & $\gamma$ & \multicolumn{4}{c}{$\frac53$} \\
  
  \hline

 \end{tabular}
\end{table*}

\subsection{Assumption of collisionality} \label{sec:knud}

\begin{figure*}
\centering
\includegraphics[width=\textwidth]{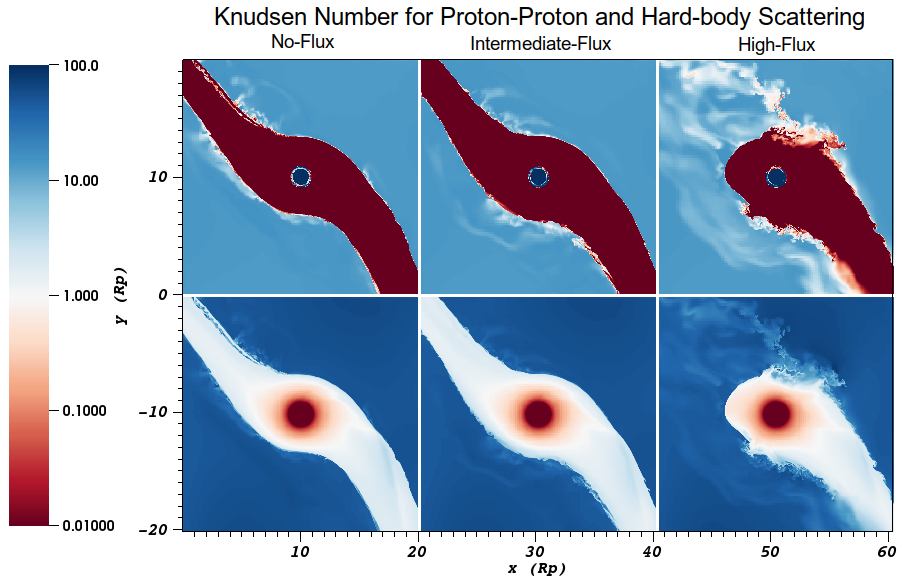}
\caption{Plots of the Knudsen number for the final states of our no-flux, intermediate-flux, and high-flux simulations. The top row shows the Knudsen number when the collisional cross-section is assumed to be due to Coulomb collisions, while the bottom row shows the Knudsen number when the cross-section is assumed to be due to hard-body collisions. The fluid is collisional throughout the entire planetary wind, with $Kn \lesssim 1$.}
\label{fig:knudsen}
\end{figure*}

As AstroBEAR is a hydrodynamics code, we implicitly assume that the fluid is collisional throughout the simulation domain, and can therefore be treated hydrodynamically rather than kinetically. We check this assumption by plotting the Knudsen number, $Kn = \lambda/L$, for the final states of the no-flux, intermediate-flux, and high-flux simulations, as shown in figure \ref{fig:knudsen}. A gas is collisional for $Kn \lesssim 1$. Here the mean free path $\lambda = (\sigma_{col} n)^{-1}$ and the characteristic length scale $L = 0.1 R_p$. We plot both the Knudsen number due to proton-proton (Coulomb) collisions (top row), where $\sigma_{col} = 10^{-13} (10^4/T)^2$ and the number density $n = \nhii$, and the Knudsen number due to neutral-proton, neutral-neutral, and proton-proton collisions (bottom row), all assumed to be hard-body collisions where $\sigma_{col} = 3.53\times10^{-16}$ and the number density $n = \nH$.

These plots show that the entire planetary wind is collisional, as also found in e.g. \citet{salz16}, save for a small portion of the blown-off material in the high-flux simulation. In particular, the arms of the wind are collisional primarily thanks to proton-proton interactions, while the base of the wind is collisional due to the high neutral density. The lack of collisionality in the ambient material can be disregarded, as the ambient does not affect the evolution of our simulations. The average mean free path at $2R_p$ using the hard-body assumption is $2.46\times10^8 \mbox{ cm}$, resulting in an average Knudsen number at $2R_p$ of $0.225$.

\section{Results} \label{sec:result}

\subsection{Analytic treatment} \label{sec:math}

To understand the influence of radiation pressure on the escaping planetary atmosphere, we first discuss an analytic approximation for the stellar Lyman-$\alpha$ flux required to completely unbind a torus of material, hereafter referred to as the "threshold flux". It was found in previous studies that hydrodynamic planetary winds will often form an extended torus-like structure filling a significant portion of the orbit \citep{matsakos15, carroll16, debrecht18}. We therefore take a simplification of such a torus as the basis for our analytic treatment of the effects of radiation pressure. We assume that the planet is injecting material into the torus with a mass-loss rate of $\dot{M} = 5.7\times10^9 \mbox{ g/s}$, which is that found in our initial-condition simulation with no Lyman-$\alpha$ flux. We remark that this value agrees with that of most other theoretical estimates \citep[e.g., see Table~1 of][]{kubyshkina2018}, as well as with observational estimates \citep[e.g.][]{vidalmadjar03}. In addition, we assume that the torus is optically thick to Lyman-$\alpha$ radiation, so that all stellar Lyman-$\alpha$ flux incident upon it is absorbed, maximizing the absorbed momentum.

To liberate material from the torus and form a cometary tail, it must be supplied with a $\Delta v$ of approximately the escape speed from the central star at the orbital distance, that is,
\begin{equation}
    \Delta v = \sqrt{\frac{2 G M_\star}{a}} = 2.16 \times 10^{7} \mbox{cm/s}.
\end{equation}
To liberate material as it is introduced into the torus, we require a force of
\begin{equation}
    \frac{dp_x}{dt} = \dot{M} \Delta v = \dot{M}\sqrt{\frac{2 G M_\star}{a}}.
\end{equation}
The force due to the stellar Lyman-$\alpha$ flux is
\begin{equation}
    \frac{dp_x}{dt} = F A p_\gamma,
\end{equation}
where the area on which the flux acts is the rectangular cross-section of the torus, $A = 2 \pi a H$, with $H$ being the full height of the torus and $F$ the flux. Equating these forces, we find that the threshold flux $F_{th}$ is given by
\begin{equation}
    F_{th} = \sqrt{\frac{2 G M_\star}{a}} \frac{\dot{M}}{A p_\gamma} = \sqrt{\frac{2 G M_\star}{a}} \frac{\dot{M}}{2 \pi a H p_\gamma}.
\end{equation}
With a torus height of $H \approx 6 R_p$, we find $F_{th} \approx 8\times10^{14} \mbox{ phot cm}^{-2} \mbox{ s}^{-1}$. This is a minimum threshold, as it applies when all particles of the torus acquire only up to the escape speed. Were most of the acquired momentum instead carried off by a small fraction of the torus, a significantly higher flux would be needed. It should be noted that this analysis does not treat the acceleration of small portions of the neutral torus material to the observed velocities of $\sim 10-100$ km/s.

\subsection{Simulation results}

All of our simulations are centered on the planet and carried out in the planet's orbiting frame of reference, with the orbital velocity vector $\boldsymbol\Omega$ in the $+z$ direction. Therefore, "up-orbit" is approximately in the $+y$ direction and "down-orbit" is approximately in the $-y$ direction. We first discuss the case without Lyman-$\alpha$ radiation pressure on the torus, which is the steady state from which the other cases are begun. We then summarize the intermediate-flux (Run 2 of table \ref{tab:runs}) and high-flux (Run 3 of table \ref{tab:runs}) cases. As the low-flux and intermediate-flux cases are essentially identical, we do not present the results of the low-flux case. Movies for each of these cases are available on the AstroBEAR YouTube channel\footnote{https://www.youtube.com/user/URAstroBEAR \label{ftn:movies}}. Much of the following discussion is similar to that from our previous study on the parameter space of planetary winds \citep{debrecht19}.

\subsubsection{Initial steady-state: no Lyman-$\alpha$ flux}

\begin{figure*}
\centering
\includegraphics[width=\textwidth]{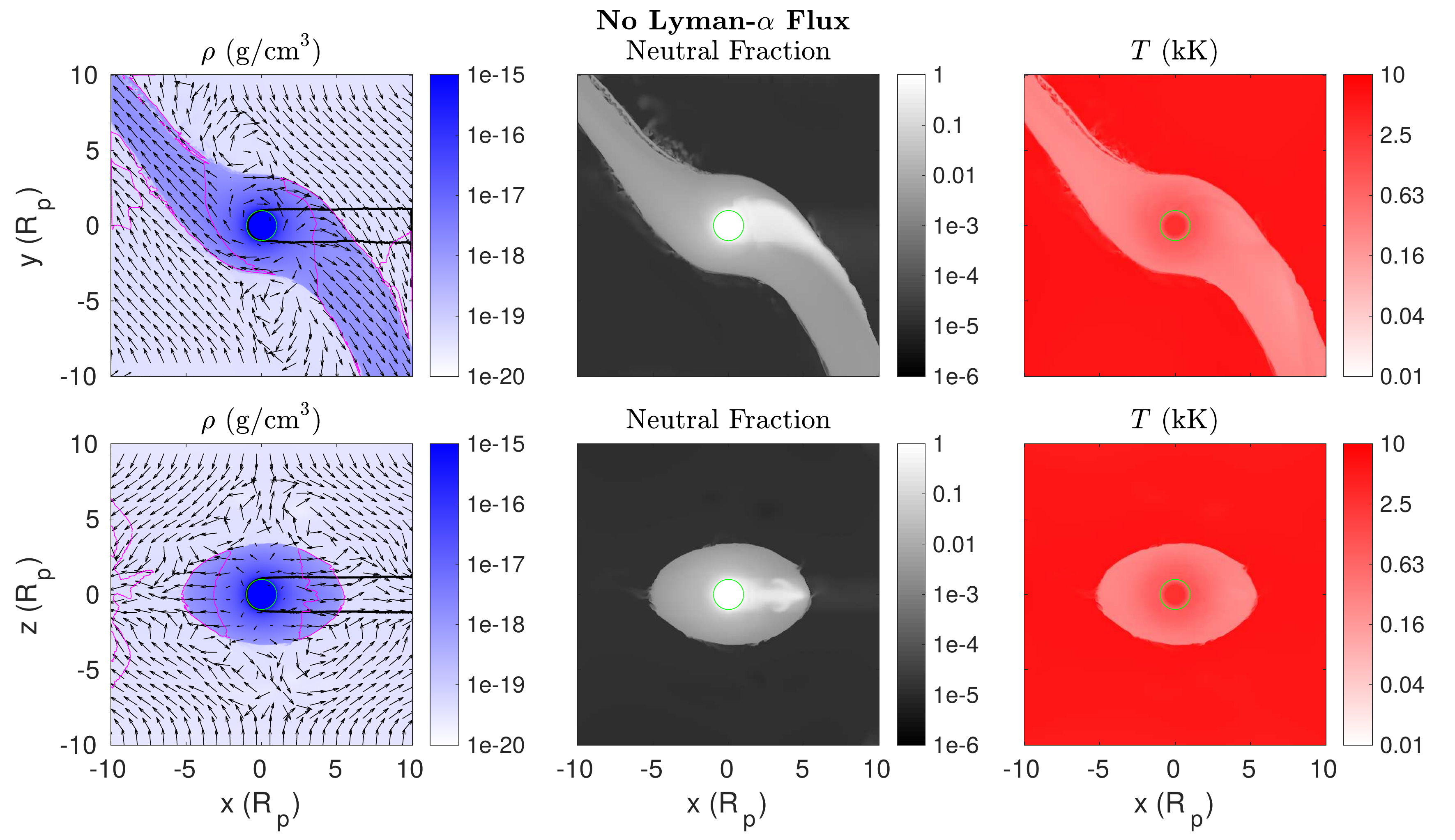}
\caption{Steady state of the initial wind, before Lyman-$\alpha$ flux is introduced. The top row shows the view looking down on the orbital plane, while the bottom row shows the view standing in the orbital plane and looking up-orbit. The left column shows density, with the magenta contour the Mach surface and the black contour the $\tau = 1$ surface and the vectors giving the velocity. The green contour is the location of the nominal planetary radius $R_p$. The center column shows the neutral fraction, and the right column shows the temperature. The star is located to the left of the simulation grid.}
\label{fig:rxt_no_flux}
\end{figure*}

The no-flux case is shown in figure \ref{fig:rxt_no_flux}. The leftmost panels of this figure show the density (hue, logarithmically scaled), in $\mbox{g cm}^{-3}$, and velocity field (quivers, normalized to the maximum velocity in the plane) for this simulation. In these panels we also show the $\tau = 1$ surface for our ionizing radiation (black contour), Mach surface (magenta contour), and nominal planetary radius $R_p$ (green contour). The simulation is carried out in the co-rotating frame. In the top row we show a 2-D slice looking down on the orbital plane (hereafter referred to as the top view). In the bottom row we show a slice in the orbital plane looking up-orbit (hereafter referred to as the side view).

The $\tau = 1$ surface denotes the position from which the wind is launched. At this point, most of the ionizing photons have been absorbed. Below the $\tau = 1$ surface the planetary hydrogen density and the corresponding optical depth quickly increase such that by $R = 0.94 R_p$, 99.9\% of the incident radiation has been absorbed. \citet{murrayclay09} found that the wind solution is insensitive to conditions below the $\tau = 1$ surface. Although the details of the flow below $\tau = 1$ are not expected to accurately model conditions of a real giant planet, this region still plays a role in the simulation by providing a flow of neutral material to larger radii that is subsequently ionized and continually supplies the wind.

The $\tau = 1$ surface also denotes the extent of the ionization shadow of the planet. The center panels of figure \ref{fig:rxt_no_flux} show the neutral hydrogen fraction (hue, logarithmically scaled) for the same slices through the planet. Here we see that the planet's ionization shadow results in material leaving the night side and remaining neutral over a significant distance. This distance is defined by the ionization timescale and the wind velocity. Also seen in this panel is that the rest of the planetary wind carries a small fraction of neutrals with it. The neutral fraction $\nhi/\nH$ is of order $10^{-2}$ in the bulk of the wind. 

The right panels in figure \ref{fig:rxt_no_flux} show the temperature, in $\mbox{K}$, on a logarithmic scale. Because the wind is mostly transparent to ionizing radiation outside of the planet's atmosphere, the temperature of the wind is determined hydrodynamically, beginning at $T = 8200 \mbox{ K}$ at the base of the wind and cooling primarily by expansion (though radiative and recombination cooling are still present). At the surface of the planet, slight asymmetry can be seen, highlighting the effect of the ionizing radiation on the equilibrium structure of the wind.

As in \citet{carroll16}, \citet{matsakos15}, and \citet{debrecht19}, we see the formation of up- and down-orbit wind trajectories. Material otherwise bound by the planet's Hill sphere is driven into a wind by pressure forces (from ionization heating) and tidal forces. Rotational forces then turn this material into the two arms of the wind. Because of the rotational forces, the wind does not fill the whole computational domain. Rather, it is confined to a torus with a quasi-cylindrical cross section, as shown in the bottom row.

\begin{figure*}
\centering
\includegraphics[width=\textwidth]{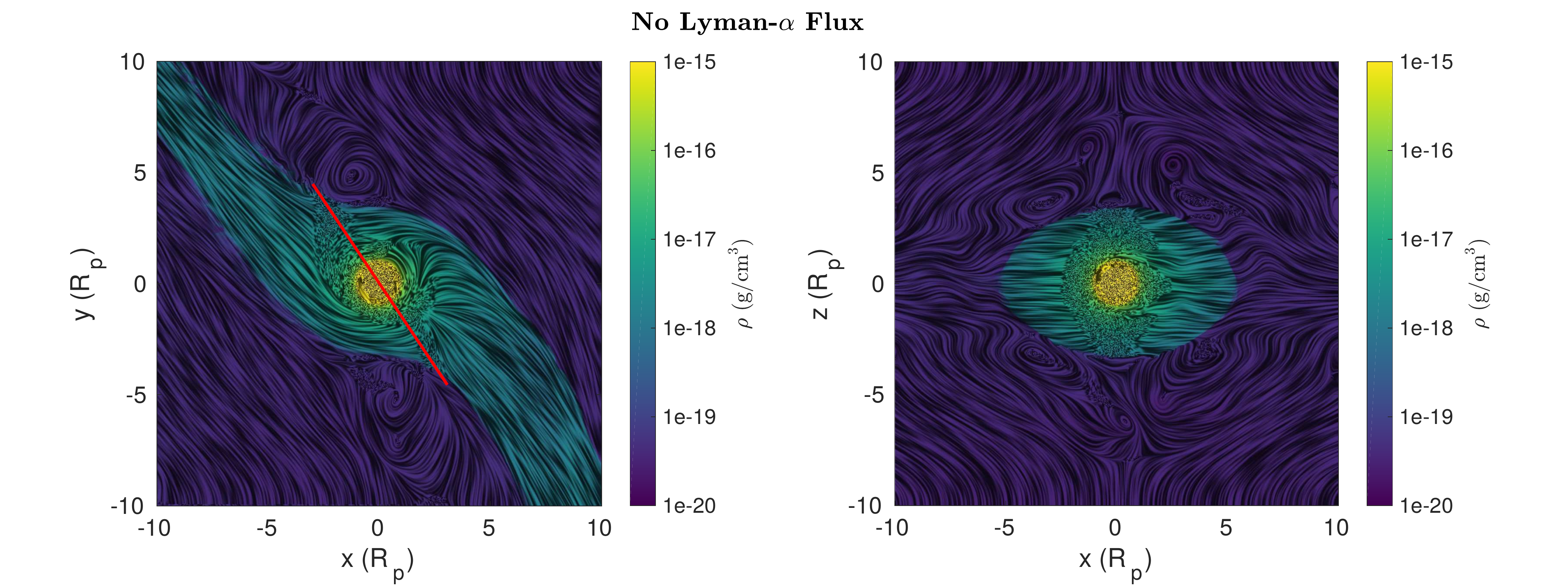}
\caption{Flow texture plot of the initial wind steady state, with the left panel showing the view looking down on the orbital plane and the right panel showing the view standing in the orbital plane looking up-orbit. The hue represents density, and the texture represents velocity streamlines. The red line divides material that forms the up-orbit arm, to the left, from material that forms the down-orbit arm, to the right. The star is located to the left of the simulation grid.}
\label{fig:flow_texture_no_flux}
\end{figure*}

We now consider the flow pattern in the model via figure \ref{fig:flow_texture_no_flux}, created by convolving random noise integrated along the streamlines of the velocity field with a color plot of the density. In this figure, hue shows density (as in the left panels of figure \ref{fig:rxt_no_flux}) and the texture of the plot represents flow streamlines. The left panel shows the top view, while the right panel shows the side view. In the top view, we can see the origin of the material in each of the arms of the wind. The up-orbit arm originates from gas in the Hill sphere to the left of the red line, while the down-orbit arm originates from gas to the right of the red line. In addition, vortices formed at the points of divergence of the arms are apparent from this plot.

The side view again shows vortices at the corners of the wind. In addition, material flowing laterally around the planet after being heated can be seen joining the bulk of the wind.

\subsubsection{Comparison with previous studies} \label{sec:cptoprev}

The steady state of this wind differs significantly from that seen in \citet{debrecht19}, with much of the difference due to the significantly denser planet. Movies of the no-flux simulation (see footnote \ref{ftn:movies}) show that the up-orbit and down-orbit arms form slowly, with material that is peeled out of the Hill sphere of the planet by tidal forces. In contrast with the previous study, these arms are not significantly redirected by the Coriolis force after formation, so that there is little complex structure to the planetary wind. The neutral tail, composed of material pulled from the night side of the planet, still forms as it did in the simulations of \citet{debrecht19}. But as the planetary wind is not turned completely around the planet by Coriolis forces the tail is not redirected by any shocks and merely flows along with the down-orbit arm. Thus it continues to exist farther away (in the $-y$ direction) from the planet than previously seen. Finally, the Mach surfaces of the wind form at a significant distance from the ionization front in the simulation of HD~209458b, while in the previous study the ionization front and Mach surface were nearly identical.

It is also interesting to compare our simulation with that of \citet{cherenkov18}, particularly their figure 1. In broad strokes, their flow appears similar to ours, with up-orbit and down-orbit arms formed by material escaping the Roche lobe of the planet. There are some subtle differences, perhaps the most significant of which is that their mass loss is much more focused through the L1 and L2 points than ours. In the simulations shown here the mass loss appears much broader in terms of the planetary surface area (though still averaging to the L1 and L2 points). The remaining differences are likely due to their inclusion of the stellar wind, which appears to force the up-orbit arm back and results in turbulence that somewhat disrupts the down-orbit arm.

\subsubsection{Intermediate Lyman-$\alpha$ flux}

\begin{figure*}
\centering
\includegraphics[width=\textwidth]{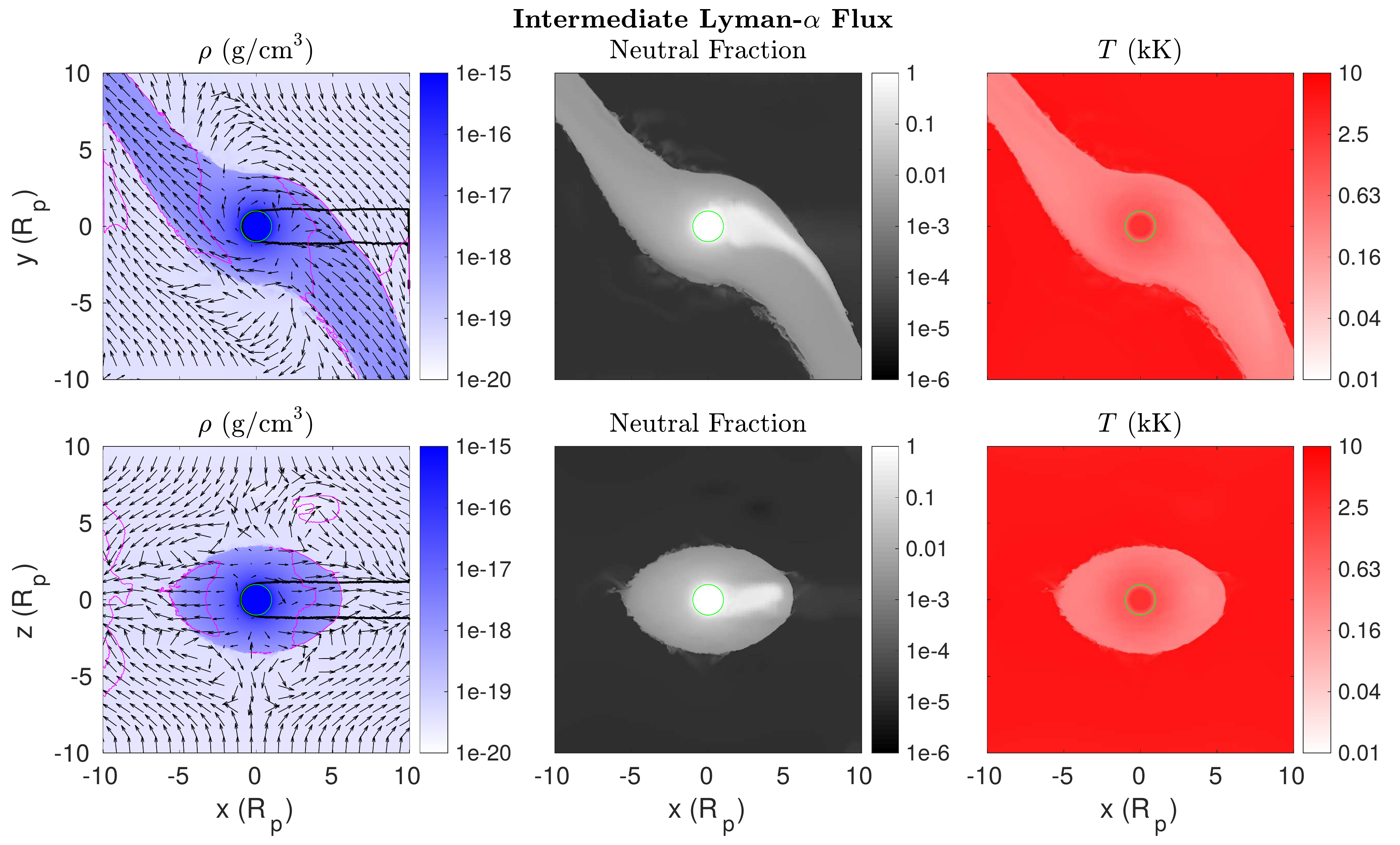}
\caption{Same as figure \ref{fig:rxt_no_flux}, for the intermediate Lyman-$\alpha$ flux case. The up- and down-orbit arms of the wind are thicker by $\sim R_p$ than in the no-flux case (Figure \ref{fig:rxt_no_flux}). The low-flux case is essentially identical to this case.}
\label{fig:rxt_med_flux}
\end{figure*}

Figure \ref{fig:rxt_med_flux} shows the steady state of the intermediate-flux run, which uses the predicted actual Lyman-$\alpha$ flux of the host star HD~209458. It can be seen that the intermediate-flux case differs only slightly from the initial steady state. The most apparent difference is that instabilities in the trailing edges of the arms of the wind are suppressed by the increase in external pressure, leading to arms that are effectively wider by approximately one planetary radius. On the other hand, instabilities along the leading edge of the wind are enhanced by the additional pressure coming from the star. However, no cometary tail forms, with both the up-orbit and down-orbit arms forming as before.

\begin{figure*}
\centering
\includegraphics[width=\textwidth]{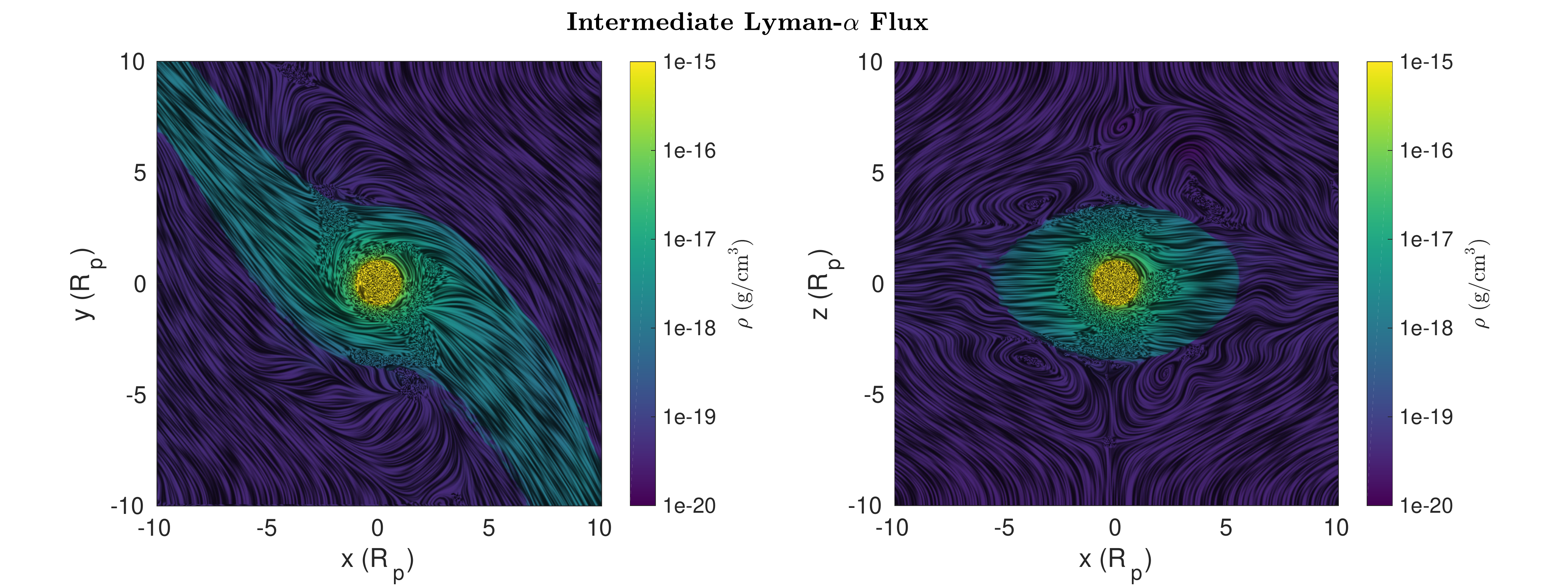}
\caption{Same as figure 2, for the intermediate Lyman-$\alpha$ flux case.}
\label{fig:flow_texture_med_flux}
\end{figure*}

The similarities between the no-flux and intermediate-flux case can also be seen in the velocity streamlines plotted in Figure \ref{fig:flow_texture_med_flux}. The origins of the up- and down-orbit arms are nearly identical to the no-flux case, with the exception that some material from near the substellar point flows completely around the planet to join the down-orbit material. In addition, the side view shows that the high-velocity portions of the wind are closer to the planet, outlining the Hill sphere more clearly in the z direction. The lateral flow around to the night side of the planet is also somewhat enhanced.

\begin{figure*}
\centering
\includegraphics[width=\textwidth]{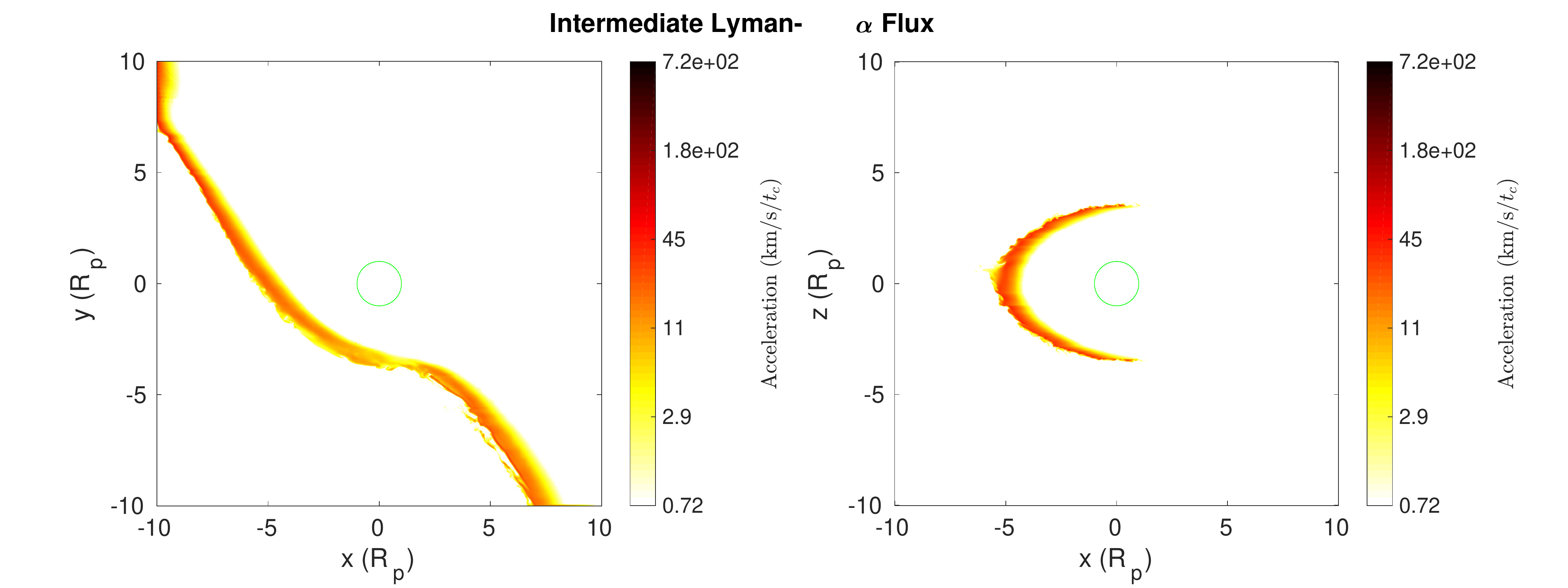}
\caption{Acceleration of wind material due to radiation pressure in the intermediate-flux case (Run 2), in units of $\mbox{km/s/}t_c$, where $t_c = 20 \mbox{ hr}$ is the crossing time of half the box (10 $R_p$). The left panel shows the top view, while the right panel shows the side view. The radiation pressure acts on a thin shell that is steadily blown over and under the rest of the wind.}
\label{fig:rad_press_med_flux}
\end{figure*}

We have plotted the acceleration of wind material due to radiation pressure in figure \ref{fig:rad_press_med_flux}, in units of km/s/$t_c$, where $t_c = 20 \mbox{ hr}$ is the crossing time for half the box (10 $R_p$). Here $0.72 \mbox{ km/s/}t_c = 1 \mbox{ cm/s}^2$. The left panel shows the top view, while the right panel shows the side view. Note that the maximum acceleration, located along the edge of the up-orbit arm around $5 R_p$, is around 100 cm/s$^2$, while the acceleration from the stellar gravity at the orbit of HD~209458b is approximately 330 cm/s$^2$. Because the wind is collisional (see section \ref{sec:knud}), this acceleration is acting not only on the material directly subject to the radiation pressure, but also on the material deeper in the wind.

\subsubsection{High Lyman-$\alpha$ flux}

The high-flux case does not reach a steady state. Instead, the up-orbit arm is pushed back periodically to join into the down-orbit arm, forming a cometary tail. We note, however, that the pressure from Lyman-$\alpha$ radiation is insufficient to completely suppress the up-orbit arm. Eventually, ionized material from the planet's Hill sphere begins to form an up-orbit arm again, which is then eventually blown back into a cometary tail. The intermittency of the cometary tail distinguishes our result from the permanent cometary tail seen in \citet{bourrier13} and \citet{Schneiter2017}.

The timescale for this cycle is the crossing time for the Hill sphere. The Hill sphere is enlarged due to the effects of radiation pressure. Modifying the standard Hill sphere derivation, we find:
\begin{equation}
    \frac{M_p}{M_\star}a^2 = \frac{r_H^2}{a}[r_H + a(\beta - 1) + 2 r_H \beta],
\end{equation}
where $\beta$ is the reduction in stellar gravity due to radiation pressure ($\beta = (g_\star - a_{rad})/g_\star$) and $r_H$ is the radius of the Hill sphere. The average size of the Hill sphere for this simulation varies, but is approximately this size of the simulation domain for much of the simulation period. The timescale for the cycling of this simulation is therefore approximately the crossing time of the simulation domain.

\begin{figure*}
\centering
\includegraphics[width=\textwidth]{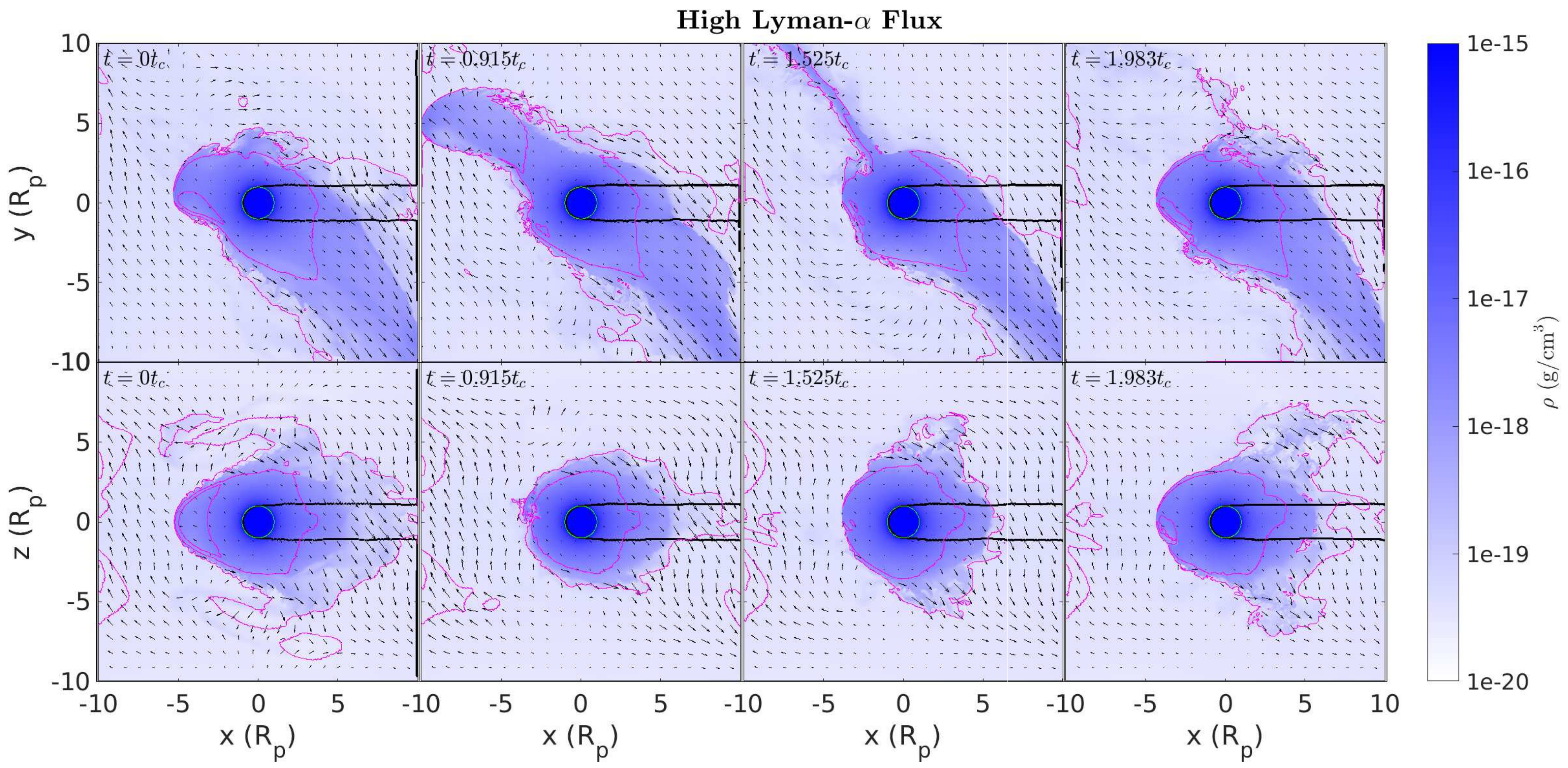}
\caption{This figure illustrates the different states of the high-flux case. The top row shows a view down onto the orbital plane and the bottom row shows the view looking up-orbit while standing in the orbital plane, with density given by hue, velocity field by quivers, Mach surface by the magenta contour, and $\tau=1$ surface by the black contour. The star is located to the left of the simulation grid. Each row shows a time series of the simulation, with panels labeled in units of the simulation crossing time $t_c = 20 \mbox{ hr}$. In the leftmost column, material begins to bubble out from the planet's Hill sphere. In the second column, the wind reaches out from the planet. In the third column, the wind material that has extended away from the planet is blown back by radiation pressure. In the final (rightmost) column, material begins to bubble out from the Hill sphere once more, and the cycle repeats itself.}
\label{fig:rho_high_flux}
\end{figure*}

Figure \ref{fig:rho_high_flux} shows the cycle using plots of the density (hue), mach number (magenta contour), velocity field (quivers), and surface of optical depth equal to one (black contour). The top row shows the top view, while the bottom row shows the side view, with time progressing from left to right. First we examine the top view. We take the left panel to be the initial state ($t = 0$ hours) for the purpose of this description. Here the radiation pressure has suppressed the up-orbit arm. Material on the up-orbit side of the planet has essentially been confined to the Hill sphere. Material that has managed to expand outwards is almost entirely subsonic.

The second panel, at $t \approx 18$ hours, shows the expansion of material from the Hill sphere to re-form a supersonic up-orbit arm. However, this arm does not extend off the grid as in the previous simulations. Instead, as we see in the third panel ($t \approx 30$ hours), as the arm expands its inertia and thermal pressure drop, allowing radiation pressure to drive it away from the star. Finally, the right panel shows the arm being blown completely backward at $t \approx 40$ hours, with most of the material flowing around the planet to join the down-orbit arm. The material on the up-orbit side is once again subsonic and inside the Hill sphere.

We now examine the side view. In the left panel, we see a relatively symmetric core of the wind with a quasi-cylindrical cross-section as in the previous simulations. We also see a large flow of diffuse material off the grid on the far side of the wind. In the second panel, the substellar portion of the wind is compressed such that its cross-section is no longer symmetric. Most of the diffuse material has blown off the grid or been recaptured by the central flow. The third panel shows material on the upper and lower edges of the wind essentially ablating due to radiation pressure. The final panel shows this material once again being driven off the grid in a diffuse stream.

\begin{figure*}
\centering
\includegraphics[width=\textwidth]{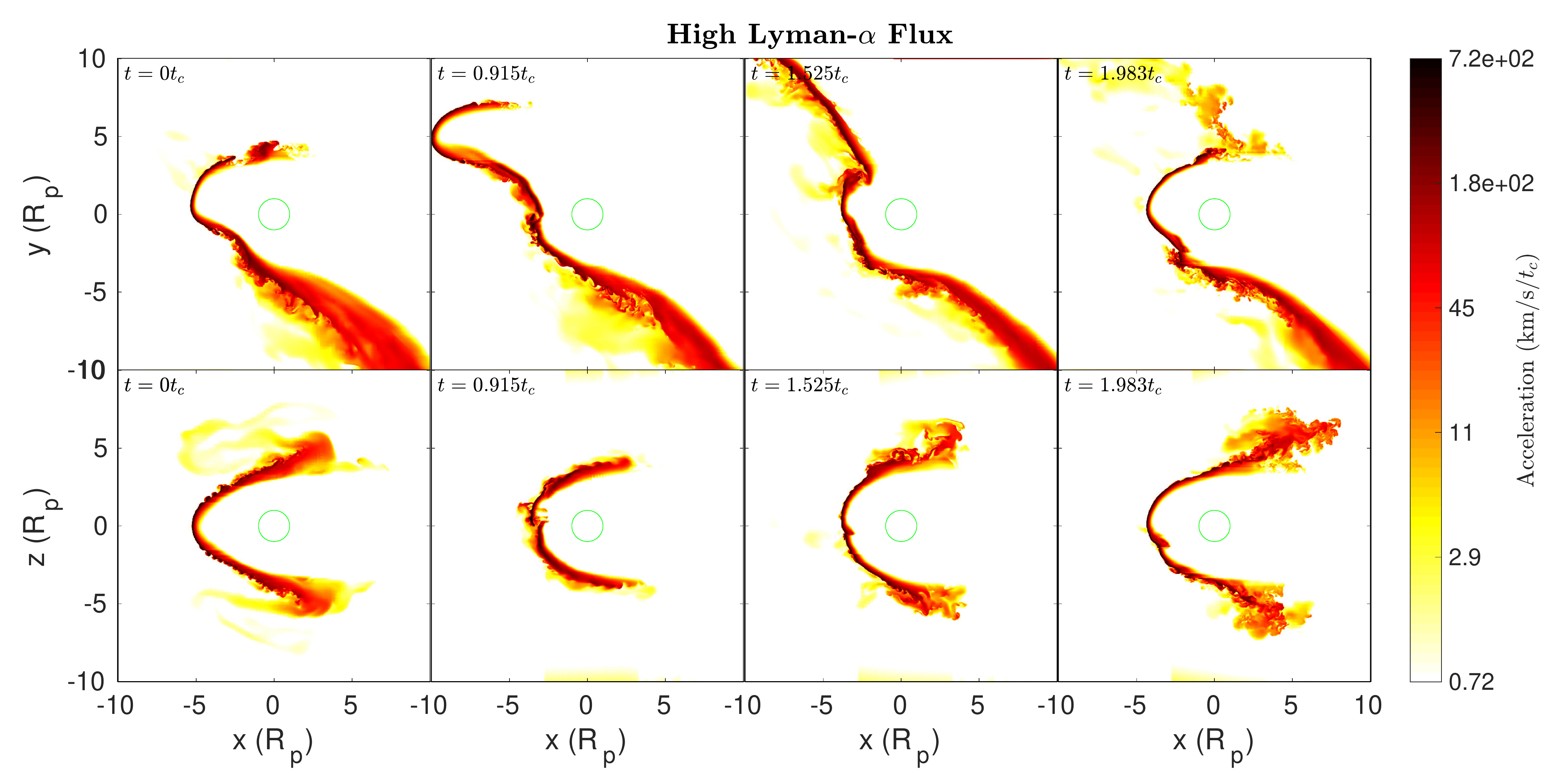}
\caption{Same as \ref{fig:rad_press_med_flux}, for the high-flux case (Run 3). Note the presence of significant absorption in the layers that can be seen in Figure \ref{fig:rho_high_flux} to be moving along the edge of the planetary wind with significant velocity.}
\label{fig:rad_press_high_flux}
\end{figure*}

As was done for the intermediate flux case, we have plotted the acceleration of the wind material due to radiation pressure in figure \ref{fig:rad_press_high_flux}. The panels are arranged as described for figure \ref{fig:rho_high_flux}. Note that in this case the maximum acceleration is greater than 1000 cm/s$^2$ along the starward edge of the wind, significantly greater than the stellar acceleration of 330 cm/s$^2$. As in the intermediate-flux case, the wind is collisional (see section \ref{sec:knud}), and the acceleration is therefore acting on material deeper in the wind in addition to the material acted upon directly by radiation pressure.

\subsection{Synthetic observations}

We also compute synthetic observations for the no-flux, intermediate-flux (Run 2), and high-flux (Run 3) cases, as described in \citet{carroll16}. Because they are done in post-processing, where computational expediency is not an issue (unlike the radiation pressure calculations, which are done on the fly), we include Doppler shifting of Lyman-$\alpha$ flux. The stellar Lyman-$\alpha$ profile used for the synthetic observations is always that taken from \citet{bourrier13}, and is not scaled with the incident flux used in the calculation of the radiation pressure. This does not affect the width of the line absorption (i.e. observed velocity), which is the principle metric of interest in the study.

\begin{figure*}
\centering
\includegraphics[width=\textwidth]{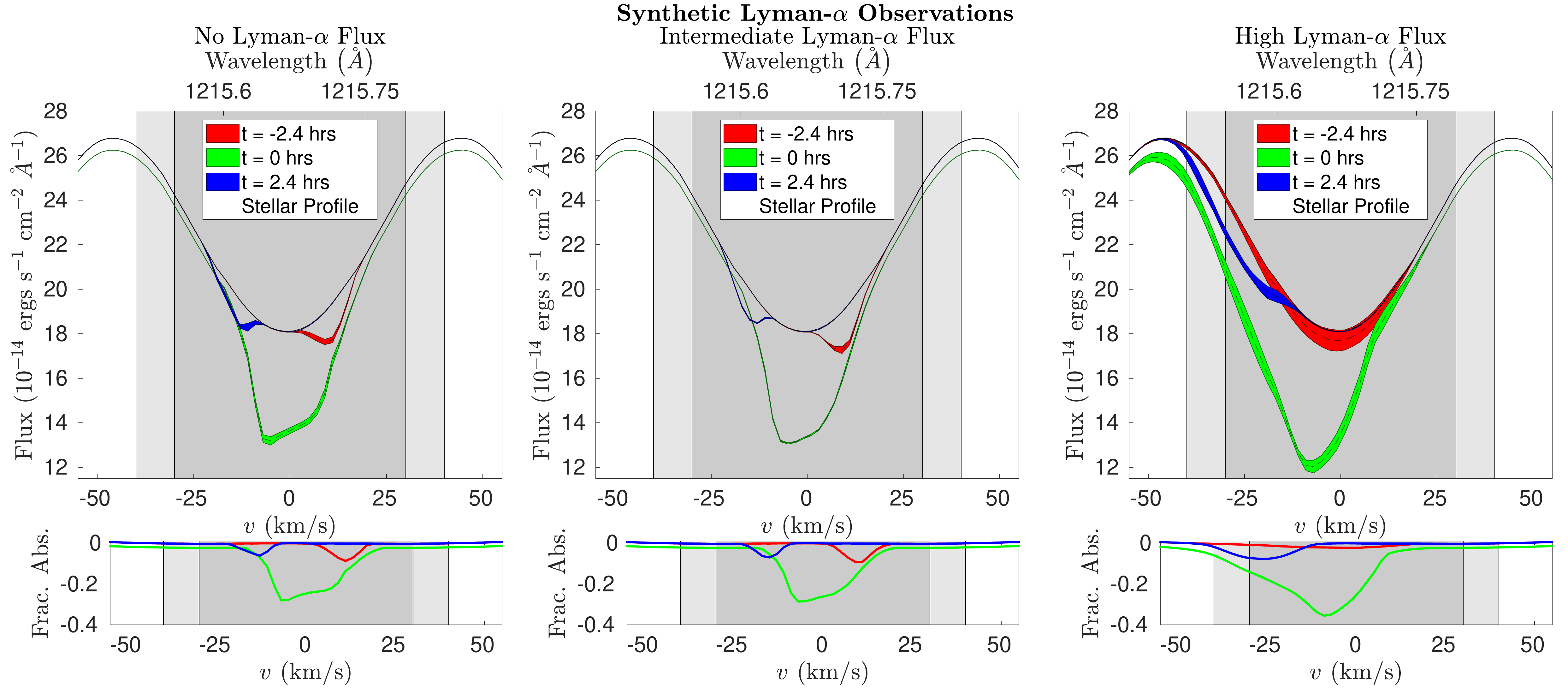}
\caption{Comparison of synthetic observations of the no-flux, intermediate-flux (Run 2), and high-flux cases (Run 3), with a focus on the core of the Lyman-$\alpha$ line. The dark grey and light grey regions represent $\pm30 \mbox{km/s}$ and $\pm40 \mbox{km/s}$, respectively, where interstellar absorption and geocoronal emission prevent the detection of planetary absorption signals. Thick lines represent observations within one standard deviation of the mean time-variable absorption, while the dashed line gives the observed average flux. The fractional absorption for each simulation is given in the bottom panels. The no-flux case and intermediate-flux case are nearly identical, with the radiation pressure serving only to reduce the instabilities in the edges of the wind and therefore decrease the variability of the simulation. The high-flux case shows additional absorption in the blue wing of the line profile as expected, in particular between -15 and -45 km/s, and greater variability in the observations; however, only a small signal is present outside of the inner region where most of the planetary absorption is located. Note that the observations at $t = 0$ always show absorption of at least 1\% due to the planetary opacity. Also note that the out-of-transit observations have artificially-lowered obscuration fractions due to the size of our simulation box.}
\label{fig:obs}
\end{figure*}

In the top panels of figure \ref{fig:obs} we show the synthetic Lyman-$\alpha$ observations for these three cases, including their variability over time. Thick lines represent observations within one standard deviation of the mean time-variable absorption, while the dashed line gives the observed average flux. Standard deviations were taken over 24.4, 18.3, and 39.66 hours for the no-flux, intermediate-flux, and high-flux cases, respectively. We see that the variability of the observations is more noticeable in the high-flux case, due to the cyclic nature of the outflow. We note that the no-flux case has a slightly larger variability than the medium-flux case. This is likely due to the previously-mentioned suppression of the edge instabilities by the external pressure.

The no-flux and intermediate-flux cases both show enhanced absorption in the red wind pre-transit and the blue wing post-transit, due to the presence of the up-orbit and down-orbit arms of the wind, respectively. Note that the line-of-sight velocity of the up-orbit arm is away from the observer, while the line-of-sight velocity of the down-orbit arm is toward the observer, while the line-of-sight component of the orbital velocity is opposite these directions. Also note the absence of enhanced absorption in the pre-transit red wing and a greater spread in velocity in the post-transit blue wing for the high-flux case, due to the periodic absence of the up-orbit arm and the acceleration of material in the down-orbit arm.

It should be noted that the out-of-transit absorptions are artificially lowered by the size of our simulation box, as there are some rays from the camera to the star that either do not or only partially pass through the simulation. 100\% of the rays pass completely through the simulation only at $\pm 17$ minutes. The out-of-transit observations should therefore be interpreted cautiously. Note however that this restriction does not alter the principle conclusion that there is no high velocity component to the outflow.

The bottom panels of figure \ref{fig:obs} show the fractional absorption of the Lyman-$\alpha$ line for each simulation. It can be seen that the absorption is confined almost entirely to the center of the line, with only the planetary transit having an effect on the wings in the lower-flux cases. The absorption differs slightly in the high-flux case, where a small proportion of excess absorption can be seen in the blue wing out to almost 50 km/s. We remark that nearly all of the absorption lies in the region covered by the interstellar medium and geocoronal emissions. This is a consequence of radiation pressure not driving sufficient material to high enough velocities.

\begin{figure*}
\includegraphics[width=\textwidth]{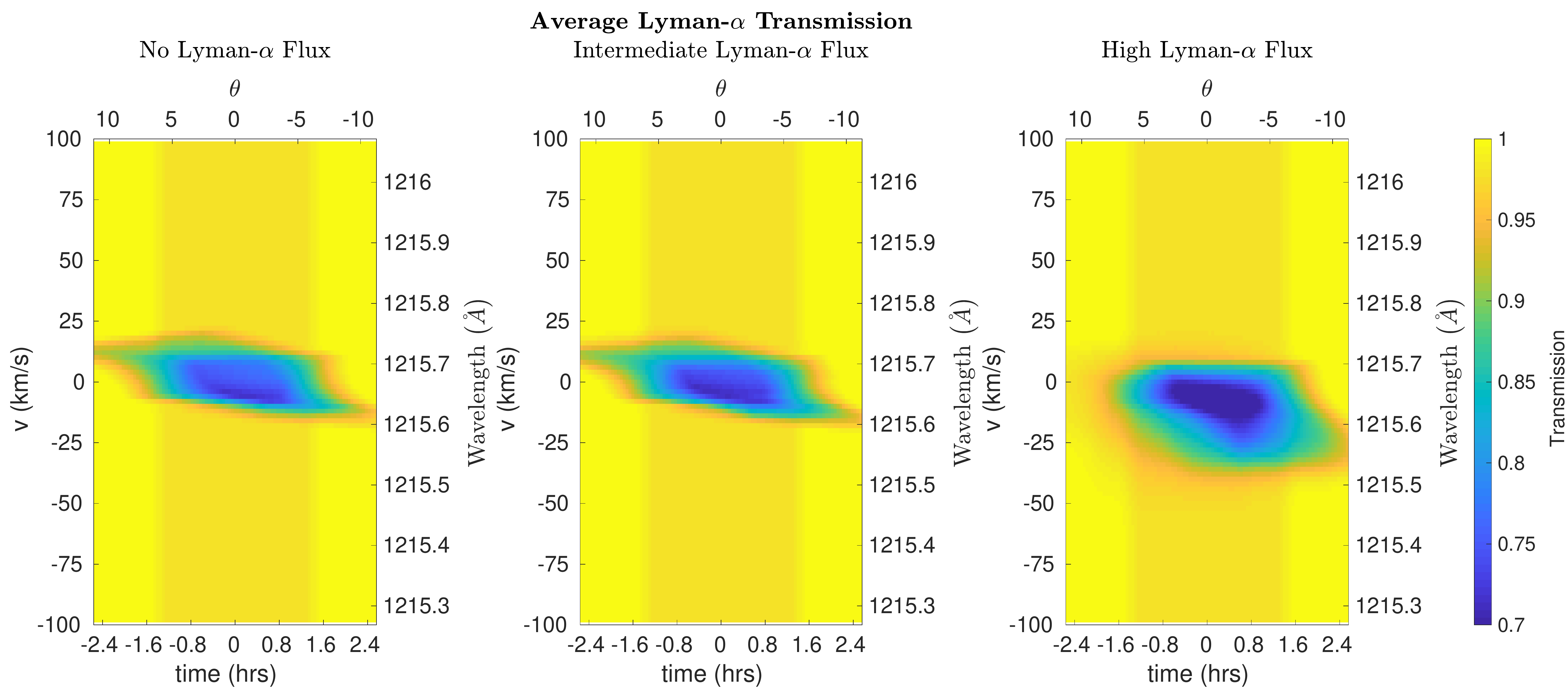}
\caption{Average transmission fraction of the Lyman-$\alpha$ flux for the no-flux, intermediate-flux (Run 2), and high-flux (Run 3) cases, as a function of time since transit/orbital angle (x axes) and line-of-sight velocity/wavelength (y axes). Again, the no-flux and intermediate-flux cases are nearly identical, with approximately 25\% absorption present only in the core of the line. The high-flux case has comparatively significant absorption near line center, with absorption of a few percent extending to approximately 45 km/s in the blue wing. As in figure \ref{fig:obs}, the out-of-transit absorption is artificially lowered due to the size of our simulation box.}
\label{fig:attenuation}
\end{figure*}

Figure \ref{fig:attenuation} is similar to Figure \ref{fig:obs}, but shows the transmission fraction as a function of time after transit (orbital angle) and velocity (wavelength). Again we see that the bulk of the absorption is confined to the center of the Lyman-$\alpha$ line, with deeper absorption extending to approximately -40 km/s in the high-flux case. Finally, we note that the Lyman-$\alpha$ absorption found in these simulations is lower than that found in \citet{carroll16}, which is due to a significantly higher ionization fraction for the bulk of the wind in the current simulations, which in turn is due to an ionization timescale in the optically-thin wind of a factor of $\sim 4$ shorter in the current simulations.

\section{Discussion}\label{sec:disc}

\subsection{Comparison with the analytic treatment}\label{sec:cptoan}

In our analytic calculation, we show that we expect a flux of $\sim 8\times10^{14} \mbox{ phot/s/cm}^2$ to be required to drive the planetary outflow away (i.e. into a cometary tail. See section \ref{sec:math}). Note that this is for a full torus of material, while in our simulations the "torus" has an angular extent of $\approx \pi/10$, with the ends of the torus outside the simulation boundaries. Estimating an actual extent of $\pi/2$, we would have a threshold flux $F_{th} = 3.2\times10^{15} \mbox{ phot/s/cm}^2$. In our simulations, we find that a flux greater than our maximum of $4.1\times10^{15} \mbox{phot/s/cm}^2$ is required to create a permanent cometary tail; however, we find that this flux, which is only $1.3\times$ the calculated threshold flux for the truncated torus, is sufficient to periodically drive the wind into a cometary tail. Thus our analytic model correctly suggests that the observed flux from HD~209458 is too low to drive cometary tails but underestimates the flux that would be needed to do so.

This underestimation may be due to two possibilities: {\it i}) the threshold flux calculated above is sufficient to blow the planetary wind away, but insufficient to completely suppress the wind; {\it ii}) the starward layer which absorbs most of the Lyman-$\alpha$ flux in our simulations is accelerated to significantly greater than escape velocity, representing an extra source of lost energy unaccounted for in the analytical calculation.

The first possibility can be thought of in terms of the ram and thermal pressure of the wind interacting with the radiation pressure from the star in two regimes: one where $P_{Ly-\alpha} \lesssim P_{th} + P_{ram}$ and one where $P_{Ly-\alpha} > P_{th} + P_{ram}$. During the period of bubbling out (right panels in figures \ref{fig:rho_high_flux} and \ref{fig:rad_press_high_flux}), the combination of the ram and thermal pressures of the planetary wind near the sonic surface is sufficient to just counter the radiation pressure from the stellar Lyman-$\alpha$ line, leading to the expansion of material and the formation of the up-orbit arm. During the period of blowback (second panels of figures \ref{fig:rho_high_flux} and \ref{fig:rad_press_high_flux}), the expansion of the wind means radiation pressure becomes greater than the sum of the planetary wind pressures and so is able to drive the up-orbit arm to larger radii.

Figures \ref{fig:rad_press_med_flux} and \ref{fig:rad_press_high_flux} show that the radiation pressure acts on a thin shell before the optical depth is sufficient to completely absorb the Lyman-$\alpha$ flux. This shell is blown around the edges of the wind while still blocking significant portions of the Lyman-$\alpha$ flux from being absorbed by the core of the outflow. This phenomenon is particularly apparent in the high-flux case (figure \ref{fig:rad_press_high_flux}). How significant this shielding is to the formation of a cometary tail can be quantified by comparing the average speed of a particle being significantly affected by radiation pressure to the escape speed $v_{esc} = 2.16 \times 10^7 \mbox{cm/s}$ of the system. In the intermediate-flux case, this average velocity is $1.6\times10^6 \mbox{ cm/s}$, while in the high-flux case, it is $3.32 \times 10^6 \mbox{ cm/s}$. It is therefore likely that material driven along the edges of the wind is a moderate, though not necessarily significant, source of energy loss not accounted for in the analytic treatment.

\begin{figure*}
\includegraphics[width=\textwidth]{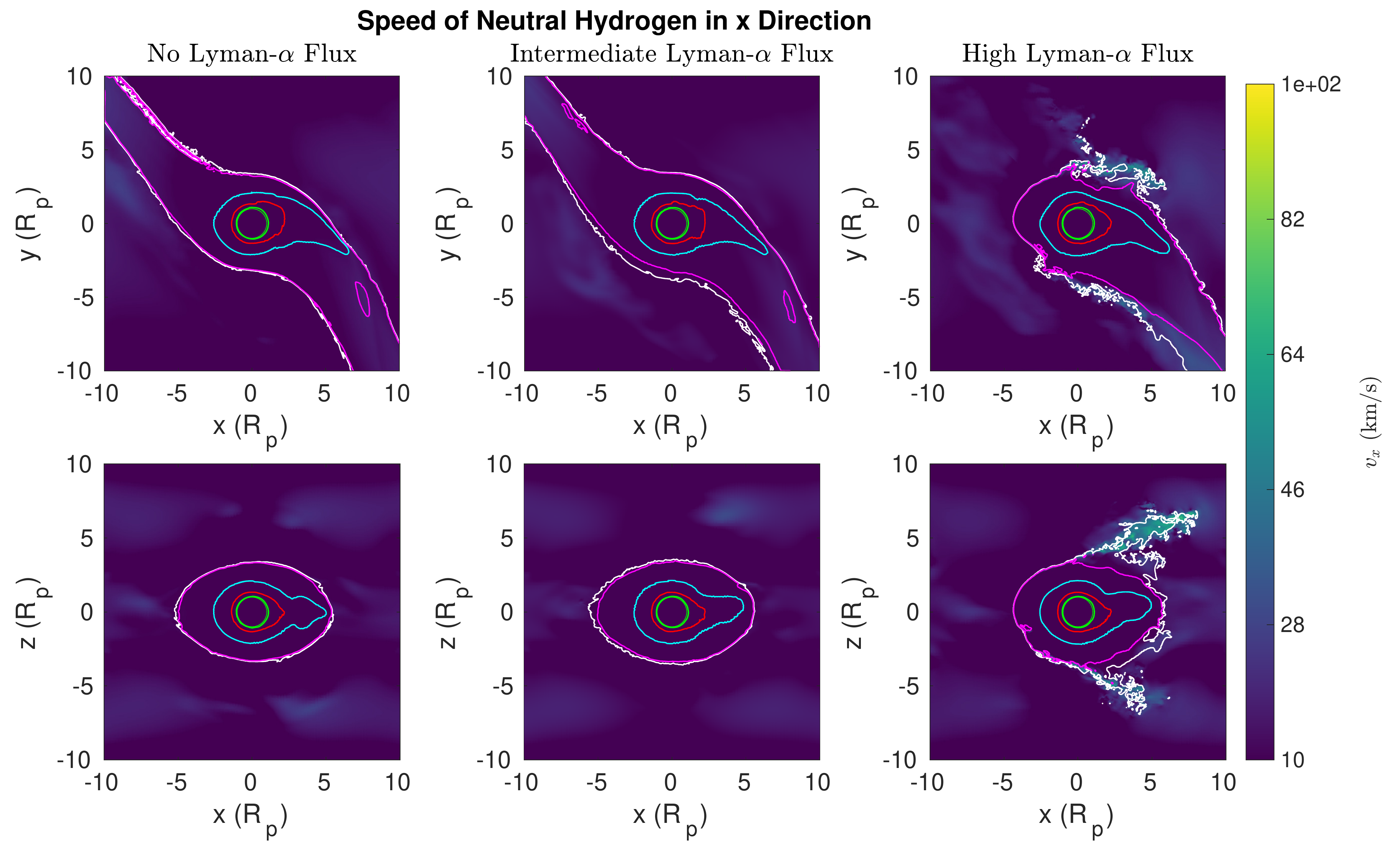}
\caption{Plots of speed with contours of density for the no (left column), intermediate (center column), and high (right column) Lyman-$\alpha$ flux simulations. The hue represents the x velocity of the gas, in km/s. The contours represent density, with green at $1.00\times10^{-16}$ g/cm$^3$, red at $3.16\times10^{-18}$ g/cm$^3$, blue at $1.00\times10^{-19}$ g/cm$^3$, magenta at $3.16\times10^{-21}$ g/cm$^3$, and white at $1.00\times10^{-22}$ g/cm$^3$. Notice that only in the high-flux case does any gas exceed 50 km/s.}
\label{fig:speed}
\end{figure*}

In addition, figure \ref{fig:speed} plots contours of density over the speed of the gas in the x direction for the no-flux, intermediate-flux, and high-flux cases. The contours are as follows: green, $1.00\times10^{-16}$ g/cm$^3$; red, $3.16\times10^{-18}$ g/cm$^3$; blue, $1.00\times10^{-19}$ g/cm$^3$; magenta, $3.16\times10^{-21}$ g/cm$^3$; white, $1.00\times10^{-22}$ g/cm$^3$. No gas exceeds a velocity of 100 km/s. The left panels show that the bulk of the up-orbit and down-orbit arms are between 10 and 50 km/s with no incident Lyman-$\alpha$ flux. The center panels indicate that the inclusion of the intermediate flux has almost no effect on the gas speed throughout the simulation. On the other hand, the right panels show that during some phases of the cycle, small amounts of neutral material are accelerated to velocities greater than 50 km/s. However, as we have seen in figure \ref{fig:obs}, the density is too low for absorption at these velocities to be significant, resulting in no absorption at the observed velocities of $\sim \pm 100$ km/s. Also note that the orbital escape speed is $\sim 200$ km/s; therefore, as discussed above, we confirm that even the high flux is insufficient to accelerate any of the wind to stellar escape speed within the space of our simulation.

Importantly, we assume in our radiation transfer that the Lyman-$\alpha$ photons are absorbed regardless of gas velocity, and thus ignore the Doppler shifts. Our calculations therefore provide an upper limit on the efficacy of radiation pressure in the formation of a cometary tail. As has been seen in figures \ref{fig:rad_press_med_flux} and \ref{fig:rad_press_high_flux}, all of the Lyman-$\alpha$ radiation is absorbed in a thin layer along the starward edge of the planetary wind. Most material along the leading edge has a speed between 10 and 50 km/s, with the majority of that in the $y$ or $z$, rather than $x$, directions. If we were to include the Doppler broadening in our absorption, this material would absorb less of the total Lyman-$\alpha$ radiation, allowing material deeper in the wind to absorb some momentum. At best, however, the same total momentum would be absorbed e.g. in the up-orbit arm, with the possibility that some significant frequency bins would have little absorption. As the creation of a cometary tail is a bulk phenomenon, only the total momentum absorbed by the up-orbit arm will have a significant influence on its formation.

Note that this does not necessarily mean that smaller amounts of material cannot be accelerated to the observed velocities of $\sim 10-100$ km/s. Indeed, if we were to include the effects of Doppler shifting in the Lyman-$\alpha$ absorption, one can imagine that material on the far edge of the wind initially moving with some small velocity in the $x$ direction could be continually accelerated by photons of increasing Doppler velocity, allowing some material to reach the observed velocities. This requires that the accelerated material not have significant density in the direction of acceleration (as we have shown that the wind is collisional) and that material between it and the star is transparent to the necessary frequencies of Lyman-$\alpha$ radiation.

\subsection{Comparison to Previous Work} \label{sec:comp}

The periodic behavior we observe in the high-flux case is qualitatively similar to the behavior found in the intermediate stellar wind regime of \citet{mccann18} (see their Figure 12). They describe the regime as being characterized by a stellar wind pressure not capable of completely confining the planetary outflow, but strong enough to disrupt the outflow once the cross-sectional area between the outflow and stellar wind becomes large enough. We note that the radiation pressure from our high-flux case ($P_{Ly-\alpha} \approx 2.2\times10^{-6} \mbox{ dyn/cm}^2$) is comparable to the pressure of \citet{mccann18}'s intermediate stellar wind ($P_\star \approx 1.3 \times10^{-6} \mbox{ dyn/cm}^2$). Therefore, it is not surprising that the same qualitative behavior is found (though the planetary outflows do differ as discussed in \ref{sec:cptoprev}). This lends credence to their suggestion that stellar radiation pressure and stellar winds should behave similarly.

It is also instructive to compare our simulations with those performed by \citet{khodachenko17}, \citet{bourrier13}, \citet{Schneiter2017}, and \citet{cherenkov18}. Our results are in good agreement with the simulations of \citet{khodachenko17}, who used a self-consistent 2-dimensional model of the wind launching due to incident EUV radiation. Those models included the effects of radiation pressure and charge exchange with the stellar wind, and found that the effects of radiation pressure on the Lyman-$\alpha$ observations were smaller than 1\% in the most favorable cases. Their EUV flux of $1.75\times10^{13}$ phot/s/cm$^2$ and Lyman-$\alpha$ flux of $4.9\times 10^{14} \mbox{ phot/s/cm}^2$ are close to those used in our simulations of the intermediate-flux case (Run 2), making them a good point of comparison. Our results are also in good agreement with the similar simulations of \citet{cherenkov18}, who resolved the planet at a moderately lower resolution than ours and the outer portions of the wind at a significantly lower resolution. They propagated their ionizing and Lyman-$\alpha$ radiation using a similar ray-tracing method as ours, and included the effects of the Doppler broadening in the Lyman-$\alpha$ line, and found that the wind envelope is essentially unaffected by Lyman-$\alpha$ flux below approximately $4 \times 10^{16} \mbox{ phot/s/cm}^2$, which supports our assertion that our flux values represent lower bounds due to our ignoring the Doppler effects.

On the other hand, both \citet{Schneiter2017} and \citet{bourrier13} find a cometary tail forming in their simulations. \citet{Schneiter2017} apply isotropic fixed boundary conditions at $3 R_p$, and use a decreased value for the stellar gravity to approximate the effects of radiation pressure. \citet{bourrier13} also use isotropic fixed boundary conditions rather than self-consistently calculated winds, with the wind launched at $2.8 R_p$ for HD~209458b. Their radiation pressure is implemented as a reduction in the gravity of the star, but is calculated with the self-shielding one would expect due to the absorption of Lyman-$\alpha$ photons in the outer layers of the wind. Perhaps most significantly, \citet{bourrier13} use a kinetic model of the planetary wind, with no pressure forces present, and remove hydrogen particles from their simulation once they've become ionized. We have seen that ionized hydrogen is a significant indirect sink for radiation pressure. Both \citet{Schneiter2017} and \citet{bourrier13} include stellar winds, and \citet{bourrier13} also include the effect of charge exchange. Thus it is difficult to perform direct comparisons to our simulations. Their results are similar to the semi-analytic models presented in \citet{carroll16}, where a small Coriolis radius and significant radiation pressure (at least 10 percent of the stellar gravity) resulted in a cometary tail. The results from \citet{Schneiter2017} and \citet{carroll16} highlight the importance of considering self-shielding when simulating the effects of radiation pressure on the evaporating winds of exoplanets. A uniform reduction in gravity is evidently insufficient to capture the full behavior of the envelope. In addition, the continued presence of the stellar wind and, in \citet{bourrier13}, charge exchange makes it difficult to disentangle which effect is primarily responsible for the cometary tail and consequently the excess absorption in the blue wing of the Lyman-$\alpha$ line.

\section{Conclusion} \label{sec:conc}

Based on these comparisons, we suggest that radiation pressure alone is insufficient to create a cometary tail in the outflow from HD~209458b. In addition, radiation pressure appears to be insufficient to create the observed absorption spectra of HD~209458b. To reproduce the observations, substantial amounts of gas must be accelerated to significantly greater than 50 km/s, which we do not see in our simulations. While the Lyman-$\alpha$ flux of HD~209458 is low in comparison to similar stars \citep{wood05}, it is not an order of magnitude lower. Thus even hot Jupiter host stars with more activity are likely to have insufficient radiation pressure to significantly affect their planets' winds. 

Both confinement by a stellar wind and charge exchange with the stellar wind are still potential sources of the fast neutral hydrogen observed in planetary winds, and a more thorough treatment of the Doppler shifting of the Lyman-$\alpha$ radiation may result in higher velocities due to radiation pressure. In fact, charge exchange has been shown in some cases to be effective in producing fast neutral hydrogen \citep[e.g.][]{ekenback10, bourrier13, bourrier16}, though \cite{bourrier13} suggest that charge exchange is not a significant contributor to those results. We will further investigate the effects of these phenomena in future studies.

\section{Acknowledgments}

We thank the Other Worlds Laboratory (OWL) at University of California, Santa Cruz for facilitating this collaboration by way of the OWL Exoplanets Summer Program, funded by the Heising-Simons Foundation. This work used the computational and visualization resources in the Center for Integrated Research Computing (CIRC) at the University of Rochester and the computational resources of the Texas Advanced Computing Center (TACC) at The University of Texas at Austin, provided through allocation TG-AST120060 from the Extreme Science and Engineering Discovery Environment (XSEDE) \citep{xsede}, which is supported by National Science Foundation grant number ACI-1548562. Financial support for this project was provided by the Department of Energy grant DE-SC0001063, the National Science Foundation grants AST-1515648, AST-1813298, and AST-1411536, and the Space Telescope Science Institute grant HST-AR-12832.01-A. EB acknowledges additional support from KITP UC Santa Barbara, funded by NSF Grant PHY-1748958, and the Aspen Center for Physics, funded by NSF Grant PHY-1607611.

\bibliography{planets.bib}

\begin{thebibliography}{}
\makeatletter
\relax
\def\mn@urlcharsother{\let\do\@makeother \do\$\do\&\do\#\do\^\do\_\do\%\do\~}
\def\mn@doi{\begingroup\mn@urlcharsother \@ifnextchar [ {\mn@doi@}
  {\mn@doi@[]}}
\def\mn@doi@[#1]#2{\def\@tempa{#1}\ifx\@tempa\@empty \href
  {http://dx.doi.org/#2} {doi:#2}\else \href {http://dx.doi.org/#2} {#1}\fi
  \endgroup}
\def\mn@eprint#1#2{\mn@eprint@#1:#2::\@nil}
\def\mn@eprint@arXiv#1{\href {http://arxiv.org/abs/#1} {{\tt arXiv:#1}}}
\def\mn@eprint@dblp#1{\href {http://dblp.uni-trier.de/rec/bibtex/#1.xml}
  {dblp:#1}}
\def\mn@eprint@#1:#2:#3:#4\@nil{\def\@tempa {#1}\def\@tempb {#2}\def\@tempc
  {#3}\ifx \@tempc \@empty \let \@tempc \@tempb \let \@tempb \@tempa \fi \ifx
  \@tempb \@empty \def\@tempb {arXiv}\fi \@ifundefined
  {mn@eprint@\@tempb}{\@tempb:\@tempc}{\expandafter \expandafter \csname
  mn@eprint@\@tempb\endcsname \expandafter{\@tempc}}}

\bibitem[\protect\citeauthoryear{{Bisikalo} \& {Cherenkov}}{{Bisikalo} \&
  {Cherenkov}}{2016}]{bisikalo16}
{Bisikalo} D.~V.,  {Cherenkov} A.~A.,  2016, \mn@doi [\azh]
  {10.1134/S1063772916020013}, \href
  {http://adsabs.harvard.edu/abs/2016ARep...60..183B} {60, 183}

\bibitem[\protect\citeauthoryear{{Bisikalo}, {Shematovich}, {Cherenkov},
  {Fossati}  \& {M{\"o}stl}}{{Bisikalo} et~al.}{2018}]{bisikalo2018}
{Bisikalo} D.~V.,  {Shematovich} V.~I.,  {Cherenkov} A.~A.,  {Fossati} L.,
  {M{\"o}stl} C.,  2018, \mn@doi [\apj] {10.3847/1538-4357/aaed21}, \href
  {http://adsabs.harvard.edu/abs/2018ApJ...869..108B} {869, 108}

\bibitem[\protect\citeauthoryear{{Bourrier} \& {Lecavelier des
  Etangs}}{{Bourrier} \& {Lecavelier des Etangs}}{2013}]{bourrier13}
{Bourrier} V.,  {Lecavelier des Etangs} A.,  2013, \mn@doi [\aap]
  {10.1051/0004-6361/201321551}, 557, A124

\bibitem[\protect\citeauthoryear{{Bourrier} et~al.,}{{Bourrier}
  et~al.}{2013}]{bourrier13b}
{Bourrier} V.,  et~al., 2013, \mn@doi [\aap] {10.1051/0004-6361/201220533},
  \href {https://ui.adsabs.harvard.edu/#abs/2013A&A...551A..63B} {551, A63}

\bibitem[\protect\citeauthoryear{{Bourrier}, {Lecavelier des Etangs},
  {Ehrenreich}, {Tanaka}  \& {Vidotto}}{{Bourrier} et~al.}{2016}]{bourrier16}
{Bourrier} V.,  {Lecavelier des Etangs} A.,  {Ehrenreich} D.,  {Tanaka} Y.~A.,
   {Vidotto} A.~A.,  2016, \mn@doi [\aap] {10.1051/0004-6361/201628362}, 591,
  A121

\bibitem[\protect\citeauthoryear{{Bourrier} et~al.,}{{Bourrier}
  et~al.}{2018}]{bourrier18}
{Bourrier} V.,  et~al., 2018, \mn@doi [\aap] {10.1051/0004-6361/201833675},
  \href {http://adsabs.harvard.edu/abs/2018A%26A...620A.147B} {620, A147}

\bibitem[\protect\citeauthoryear{Carroll-Nellenback, Shroyer, Frank  \&
  Ding}{Carroll-Nellenback et~al.}{2013}]{carroll13}
Carroll-Nellenback J.~J.,  Shroyer B.,  Frank A.,   Ding C.,  2013, \mn@doi
  [Journal of Computational Physics]
  {http://dx.doi.org/10.1016/j.jcp.2012.10.004}, 236, 461

\bibitem[\protect\citeauthoryear{{Carroll-Nellenback}, {Frank}, {Liu},
  {Quillen}, {Blackman}  \& {Dobbs-Dixon}}{{Carroll-Nellenback}
  et~al.}{2017}]{carroll16}
{Carroll-Nellenback} J.,  {Frank} A.,  {Liu} B.,  {Quillen} A.~C.,  {Blackman}
  E.~G.,   {Dobbs-Dixon} I.,  2017, \mn@doi [MNRAS] {10.1093/mnras/stw3307},
  466, 2458

\bibitem[\protect\citeauthoryear{{Cherenkov}, {Bisikalo}, {Fossati}  \&
  {M{\"o}stl}}{{Cherenkov} et~al.}{2017}]{cherenkov2017}
{Cherenkov} A.,  {Bisikalo} D.,  {Fossati} L.,   {M{\"o}stl} C.,  2017, \mn@doi
  [\apj] {10.3847/1538-4357/aa82b2}, \href
  {http://adsabs.harvard.edu/abs/2017ApJ...846...31C} {846, 31}

\bibitem[\protect\citeauthoryear{{Cherenkov}, {Bisikalo}  \&
  {Kosovichev}}{{Cherenkov} et~al.}{2018}]{cherenkov18}
{Cherenkov} A.~A.,  {Bisikalo} D.~V.,   {Kosovichev} A.~G.,  2018, \mn@doi
  [\mnras] {doi:10.1093/mnras/stx3230}, \href
  {http://adsabs.harvard.edu/abs/2018MNRAS.475..605C} {475, 605}

\bibitem[\protect\citeauthoryear{{Christie}, {Arras}  \& {Li}}{{Christie}
  et~al.}{2016}]{christie16}
{Christie} D.,  {Arras} P.,   {Li} Z.,  2016, \mn@doi [\apj]
  {10.3847/0004-637X/820/1/3}, 820, 3

\bibitem[\protect\citeauthoryear{{Cunningham}, {Frank}, {Varni{\`e}re},
  {Mitran}  \& {Jones}}{{Cunningham} et~al.}{2009}]{cunningham09}
{Cunningham} A.~J.,  {Frank} A.,  {Varni{\`e}re} P.,  {Mitran} S.,   {Jones}
  T.~W.,  2009, \mn@doi [\apjs] {10.1088/0067-0049/182/2/519}, \href
  {http://adsabs.harvard.edu/abs/2009ApJS..182..519C} {182, 519}

\bibitem[\protect\citeauthoryear{{Debrecht}, {Carroll-Nellenback}, {Frank},
  {Fossati}, {Blackman}  \& {Dobbs-Dixon}}{{Debrecht}
  et~al.}{2018}]{debrecht18}
{Debrecht} A.,  {Carroll-Nellenback} J.,  {Frank} A.,  {Fossati} L.,
  {Blackman} E.~G.,   {Dobbs-Dixon} I.,  2018, \mn@doi [\mnras]
  {10.1093/mnras/sty1164}, 478, 2592

\bibitem[\protect\citeauthoryear{{Debrecht}, {Carroll-Nellenback}, {Frank},
  {McCann}, {Murray-Clay}  \& {Blackman}}{{Debrecht} et~al.}{2019}]{debrecht19}
{Debrecht} A.,  {Carroll-Nellenback} J.,  {Frank} A.,  {McCann} J.,
  {Murray-Clay} R.,   {Blackman} E.~G.,  2019, \mn@doi [\mnras]
  {10.1093/mnras/sty3212}, 483, 1481

\bibitem[\protect\citeauthoryear{{Ehrenreich} et~al.,}{{Ehrenreich}
  et~al.}{2015}]{ehrenreich15}
{Ehrenreich} D.,  et~al., 2015, \mn@doi [\nat] {10.1038/nature14501}, \href
  {https://ui.adsabs.harvard.edu/#abs/2015Natur.522..459E} {522, 459}

\bibitem[\protect\citeauthoryear{{Ekenb{\"a}ck}, {Holmstr{\"o}m}, {Wurz},
  {Grie{\ss}meier}, {Lammer}, {Selsis}  \& {Penz}}{{Ekenb{\"a}ck}
  et~al.}{2010}]{ekenback10}
{Ekenb{\"a}ck} A.,  {Holmstr{\"o}m} M.,  {Wurz} P.,  {Grie{\ss}meier} J.~M.,
  {Lammer} H.,  {Selsis} F.,   {Penz} T.,  2010, \mn@doi [\apj]
  {10.1088/0004-637X/709/2/670}, \href
  {https://ui.adsabs.harvard.edu/abs/2010ApJ...709..670E} {709, 670}

\bibitem[\protect\citeauthoryear{{Fossati} et~al.,}{{Fossati}
  et~al.}{2010}]{fossati10a}
{Fossati} L.,  et~al., 2010, \mn@doi [\apjl] {10.1088/2041-8205/714/2/L222},
  \href {https://ui.adsabs.harvard.edu/abs/2010ApJ...714L.222F} {714, L222}

\bibitem[\protect\citeauthoryear{{Garc{\'i}a Mu{\~n}oz}}{{Garc{\'i}a
  Mu{\~n}oz}}{2007}]{garciamunoz07}
{Garc{\'i}a Mu{\~n}oz} A.,  2007, \mn@doi [\planss]
  {10.1016/j.pss.2007.03.007}, 55, 1426

\bibitem[\protect\citeauthoryear{{Holmstr{\"o}m}, {Ekenb{\"a}ck}, {Selsis},
  {Penz}, {Lammer}  \& {Wurz}}{{Holmstr{\"o}m} et~al.}{2008}]{holmstrom2008}
{Holmstr{\"o}m} M.,  {Ekenb{\"a}ck} A.,  {Selsis} F.,  {Penz} T.,  {Lammer} H.,
    {Wurz} P.,  2008, \mn@doi [\nat] {10.1038/nature06600}, \href
  {http://adsabs.harvard.edu/abs/2008Natur.451..970H} {451, 970}

\bibitem[\protect\citeauthoryear{{Khodachenko}, {Shaikhislamov}, {Lammer}  \&
  {Prokopov}}{{Khodachenko} et~al.}{2015}]{khodachenko15}
{Khodachenko} M.~L.,  {Shaikhislamov} I.~F.,  {Lammer} H.,   {Prokopov} P.~A.,
  2015, \mn@doi [\apj] {10.1088/0004-637X/813/1/50}, 813, 50

\bibitem[\protect\citeauthoryear{{Khodachenko} et~al.,}{{Khodachenko}
  et~al.}{2017}]{khodachenko17}
{Khodachenko} M.~L.,  et~al., 2017, \mn@doi [\apj] {10.3847/1538-4357/aa88ad},
  847, 126

\bibitem[\protect\citeauthoryear{{Kislyakova}, {Holmstr{\"o}m}, {Lammer},
  {Odert}  \& {Khodachenko}}{{Kislyakova} et~al.}{2014}]{kislyakova2014}
{Kislyakova} K.~G.,  {Holmstr{\"o}m} M.,  {Lammer} H.,  {Odert} P.,
  {Khodachenko} M.~L.,  2014, \mn@doi [Science] {10.1126/science.1257829},
  \href {http://adsabs.harvard.edu/abs/2014Sci...346..981K} {346, 981}

\bibitem[\protect\citeauthoryear{{Koskinen}, {Harris}, {Yelle}  \&
  {Lavvas}}{{Koskinen} et~al.}{2013}]{koskinen2013a}
{Koskinen} T.~T.,  {Harris} M.~J.,  {Yelle} R.~V.,   {Lavvas} P.,  2013,
  \mn@doi [\icarus] {10.1016/j.icarus.2012.09.027}, \href
  {http://adsabs.harvard.edu/abs/2013Icar..226.1678K} {226, 1678}

\bibitem[\protect\citeauthoryear{{Krumholz}, {Stone}  \& {Gardiner}}{{Krumholz}
  et~al.}{2007}]{krumholz07}
{Krumholz} M.~R.,  {Stone} J.~M.,   {Gardiner} T.~A.,  2007, \mn@doi [\apj]
  {10.1086/522665}, 671, 518

\bibitem[\protect\citeauthoryear{{Kubyshkina} et~al.,}{{Kubyshkina}
  et~al.}{2018}]{kubyshkina2018}
{Kubyshkina} D.,  et~al., 2018, \mn@doi [\aap] {10.1051/0004-6361/201833737},
  \href {http://ukads.nottingham.ac.uk/abs/2018A%26A...619A.151K} {619, A151}

\bibitem[\protect\citeauthoryear{{Kulow}, {France}, {Linsky}  \& {Parke
  Loyd}}{{Kulow} et~al.}{2014}]{kulow14}
{Kulow} J.~R.,  {France} K.,  {Linsky} J.,   {Parke Loyd} R.~O.,  2014, \mn@doi
  [\apj] {10.1088/0004-637X/786/2/132}, \href
  {https://ui.adsabs.harvard.edu/#abs/2014ApJ...786..132K/} {786, 132}

\bibitem[\protect\citeauthoryear{{Lammer}, {Selsis}, {Ribas}, {Guinan}, {Bauer}
   \& {Weiss}}{{Lammer} et~al.}{2003}]{lammer2003}
{Lammer} H.,  {Selsis} F.,  {Ribas} I.,  {Guinan} E.~F.,  {Bauer} S.~J.,
  {Weiss} W.~W.,  2003, \mn@doi [\apjl] {10.1086/380815}, \href
  {http://adsabs.harvard.edu/abs/2003ApJ...598L.121L} {598, L121}

\bibitem[\protect\citeauthoryear{{Lecavelier Des Etangs}, {Vidal-Madjar}  \&
  {Desert}}{{Lecavelier Des Etangs} et~al.}{2008}]{lecavelier08}
{Lecavelier Des Etangs} A.,  {Vidal-Madjar} A.,   {Desert} J.~M.,  2008,
  \mn@doi [\nat] {10.1038/nature07402}, \href
  {https://ui.adsabs.harvard.edu/abs/2008Natur.456E...1L} {456, E1}

\bibitem[\protect\citeauthoryear{{Lecavelier des Etangs}, {Vidal-Madjar},
  {McConnell}  \& {H{\'e}brard}}{{Lecavelier des Etangs}
  et~al.}{2004}]{lecavelier04}
{Lecavelier des Etangs} A.,  {Vidal-Madjar} A.,  {McConnell} J.~C.,
  {H{\'e}brard} G.,  2004, \mn@doi [\aap] {10.1051/0004-6361:20040106}, \href
  {https://ui.adsabs.harvard.edu/abs/2004A&A...418L...1L} {418, L1}

\bibitem[\protect\citeauthoryear{{Lecavelier des Etangs} et~al.,}{{Lecavelier
  des Etangs} et~al.}{2010}]{lecavelier10}
{Lecavelier des Etangs} A.,  et~al., 2010, \mn@doi [\aap]
  {10.1051/0004-6361/200913347}, 514

\bibitem[\protect\citeauthoryear{{Lecavelier des Etangs} et~al.,}{{Lecavelier
  des Etangs} et~al.}{2012}]{lecavelier12}
{Lecavelier des Etangs} A.,  et~al., 2012, \mn@doi [\aap]
  {10.1051/0004-6361/201219363}, \href
  {http://cdsads.u-strasbg.fr/abs/2012A\%26A...543L...4L} {543, L4}

\bibitem[\protect\citeauthoryear{{Matsakos}, {Uribe}  \&
  {K{\"o}nigl}}{{Matsakos} et~al.}{2015}]{matsakos15}
{Matsakos} T.,  {Uribe} A.,   {K{\"o}nigl} A.,  2015, \mn@doi [\aap]
  {10.1051/0004-6361/201425593}, 578, A6

\bibitem[\protect\citeauthoryear{{McCann}, {Murray-Clay}, {Kratter}  \&
  {Krumholz}}{{McCann} et~al.}{2019}]{mccann18}
{McCann} J.,  {Murray-Clay} R.~A.,  {Kratter} K.,   {Krumholz} M.~R.,  2019,
  \mn@doi [\apj] {10.3847/1538-4357/ab05b8}, \href
  {https://ui.adsabs.harvard.edu/abs/2019ApJ...873...89M} {873, 89}

\bibitem[\protect\citeauthoryear{{Murray-Clay}, {Chiang}  \&
  {Murray}}{{Murray-Clay} et~al.}{2009}]{murrayclay09}
{Murray-Clay} R.~A.,  {Chiang} E.~I.,   {Murray} N.,  2009, \mn@doi [\apj]
  {10.1088/0004-637X/693/1/23}, \href
  {http://adsabs.harvard.edu/abs/2009ApJ...693...23M} {693, 23}

\bibitem[\protect\citeauthoryear{{Owen} \& {Adams}}{{Owen} \&
  {Adams}}{2014}]{owen14}
{Owen} J.~E.,  {Adams} F.~C.,  2014, \mn@doi [MNRAS] {10.1093/mnras/stu1684},
  444, 3761

\bibitem[\protect\citeauthoryear{{Salz}, {Czesla}, {Schneider}  \&
  {Schmitt}}{{Salz} et~al.}{2016}]{salz16}
{Salz} M.,  {Czesla} S.,  {Schneider} P.~C.,   {Schmitt} J.~H.~M.~M.,  2016,
  \mn@doi [\aap] {10.1051/0004-6361/201526109}, \href
  {https://ui.adsabs.harvard.edu/abs/2016A&A...586A..75S} {586, A75}

\bibitem[\protect\citeauthoryear{{Schneiter}, {Vel{\'a}zquez}, {Esquivel},
  {Raga}  \& {Blanco-Cano}}{{Schneiter} et~al.}{2007}]{Schneiter2007}
{Schneiter} E.~M.,  {Vel{\'a}zquez} P.~F.,  {Esquivel} A.,  {Raga} A.~C.,
  {Blanco-Cano} X.,  2007, \mn@doi [\apjl] {10.1086/524945}, \href
  {http://adsabs.harvard.edu/abs/2007ApJ...671L..57S} {671, L57}

\bibitem[\protect\citeauthoryear{{Schneiter}, {Esquivel}, {D'Angelo},
  {Vel{\'a}zquez}, {Raga}  \& {Costa}}{{Schneiter}
  et~al.}{2016}]{Schneiter2017}
{Schneiter} E.~M.,  {Esquivel} A.,  {D'Angelo} C.~S.~V.,  {Vel{\'a}zquez}
  P.~F.,  {Raga} A.~C.,   {Costa} A.,  2016, \mn@doi [\mnras]
  {10.1093/mnras/stw076}, \href
  {http://adsabs.harvard.edu/abs/2016MNRAS.457.1666S} {457, 1666}

\bibitem[\protect\citeauthoryear{{Stassun}, {Collins}  \& {Gaudi}}{{Stassun}
  et~al.}{2017}]{stassun17}
{Stassun} K.~G.,  {Collins} K.~A.,   {Gaudi} B.~S.,  2017, \mn@doi [\aj]
  {10.3847/1538-3881/aa5df3}, \href
  {http://adsabs.harvard.edu/abs/2017AJ....153..136S} {153, 136}

\bibitem[\protect\citeauthoryear{{Teyssandier}, {Owen}, {Adams}  \&
  {Quillen}}{{Teyssandier} et~al.}{2015}]{teyssandier15}
{Teyssandier} J.,  {Owen} J.~E.,  {Adams} F.~C.,   {Quillen} A.~C.,  2015,
  \mn@doi [\mnras] {10.1093/mnras/stv1386}, 452, 1743

\bibitem[\protect\citeauthoryear{{Tian}, {Toon}, {Pavlov}  \& {De
  Sterck}}{{Tian} et~al.}{2005}]{tian05}
{Tian} F.,  {Toon} O.~B.,  {Pavlov} A.~A.,   {De Sterck} H.,  2005, \mn@doi
  [\apj] {10.1086/427204}, \href
  {https://ui.adsabs.harvard.edu/abs/2005ApJ...621.1049T} {621, 1049}

\bibitem[\protect\citeauthoryear{{Towns}, {Cockerill}, {Dahan}  \&
  {Foster}}{{Towns} et~al.}{2014}]{xsede}
{Towns} J.,  {Cockerill} T.,  {Dahan} M.,   {Foster} I.,  2014, \mn@doi
  [Computing in Science and Engineering] {10.1109/MCSE.2014.80}, 16, 62

\bibitem[\protect\citeauthoryear{{Tremblin} \& {Chiang}}{{Tremblin} \&
  {Chiang}}{2013}]{tremblin13}
{Tremblin} P.,  {Chiang} E.,  2013, \mn@doi [\mnras] {10.1093/mnras/sts212},
  428, 2565

\bibitem[\protect\citeauthoryear{{Vidal-Madjar}, {Lecavelier des Etangs},
  {D{\'e}sert}, {Ballester}, {Ferlet}, {H{\'e}brard}  \&
  {Mayor}}{{Vidal-Madjar} et~al.}{2003}]{vidalmadjar03}
{Vidal-Madjar} A.,  {Lecavelier des Etangs} A.,  {D{\'e}sert} J.-M.,
  {Ballester} G.~E.,  {Ferlet} R.,  {H{\'e}brard} G.,   {Mayor} M.,  2003,
  \mn@doi [\nat] {10.1038/nature01448}, \href
  {http://adsabs.harvard.edu/abs/2003Natur.422..143V} {422, 143}

\bibitem[\protect\citeauthoryear{{Villarreal D'Angelo}, {Esquivel}, {Schneiter}
   \& {Sgr{\'o}}}{{Villarreal D'Angelo} et~al.}{2018}]{villarreal18}
{Villarreal D'Angelo} C.,  {Esquivel} A.,  {Schneiter} M.,   {Sgr{\'o}} M.~A.,
  2018, \mn@doi [\mnras] {10.1093/mnras/sty1544}, 479, 3115

\bibitem[\protect\citeauthoryear{{Wood}, {Redfield}, {Linsky}, {M{\"u}ller}  \&
  {Zank}}{{Wood} et~al.}{2005}]{wood05}
{Wood} B.~E.,  {Redfield} S.,  {Linsky} J.~L.,  {M{\"u}ller} H.,   {Zank}
  G.~P.,  2005, \mn@doi [\apjs] {10.1086/430523}, \href
  {http://adsabs.harvard.edu/abs/2005ApJS..159..118W} {159, 118}

\bibitem[\protect\citeauthoryear{{Yelle}}{{Yelle}}{2004}]{yelle2004}
{Yelle} R.~V.,  2004, \mn@doi [\icarus] {10.1016/j.icarus.2004.02.008}, \href
  {http://adsabs.harvard.edu/abs/2004Icar..170..167Y} {170, 167}

\makeatother
\end{thebibliography}

\bsp

\label{lastpage}

\end{document}